\documentclass[amsmath,amssymb,superscriptaddress,aps,showpacs,10pt,twocolumn,nofootinbib]{revtex4-1}

\usepackage{amsfonts}
\usepackage{graphicx}
\usepackage{xcolor}
\usepackage{soul}
\usepackage{float}

\newcommand{\bfmath}[1]{\boldsymbol{#1}}

\newcommand{\grad}{\bfmath{\nabla}}

\setcitestyle{numbers,square,citesep={,\kern-.3em}}
\begin{document}

\title{Excitonic giant-dipole potentials in cuprous oxide}

\date{\today}
\author{Markus Kurz, Peter Gr\"unwald, Stefan Scheel}
\affiliation{Institut f\"ur Physik, Universit\"at Rostock, Albert-Einstein-Stra{\ss}e 23, D-18059 Rostock, Germany}

\begin{abstract}
In this work we predict the existence of a novel species of Wannier excitons when exposed to crossed electric and magnetic fields. In particular, we present a theory of giant-dipole excitons in $\textrm{Cu}_2\rm O$ in crossed fields. Within our theoretical approach we perform a pseudoseparation of the center-of-mass motion for the field-dressed excitonic species, thereby obtaining an effective single-particle Hamiltonian for the relative motion. For arbitrary gauge fields we exactly separate the gauge-dependent kinetic energy terms from the effective single-particle interaction potential. Depending on the applied field strengths and the specific field orientation, the potential for the relative motion of electron and hole exhibits an outer well at spatial separations up to several micrometers and depths up to $380\, \mu \rm eV$, leading to possible permanent excitonic electric dipole moments of around three million Debye. 
\end{abstract}
\maketitle
\section{Introduction}
Wannier excitons are of great physical interest as they are the quanta of the fundamental optical excitation in semiconductors \cite{Frenkel1931, Mott1938}. Excitons consist of a negatively charged electron in the conduction band and a positively charged hole in the valence band. As the interaction between the two species can be modeled as a screened Coulomb interaction, excitons are often considered to be a solid-state quasi-particle analogue to the hydrogen atom \cite{Gross1952,Hayashi1952,Gross1956}. In particular, excitons in cuprous oxide ($\textrm{Cu}_2\textrm{O}$) have attracted quite some attention in recent years due to an outstanding experiment, in which the hydrogen-like absorption spectrum of these quasi-particles could be observed up to principal quantum numbers $n=25$ \cite{Scheel2014}.

However, the hydrogen-like model of excitons is generally too simple to describe the spectra adequately. It has been shown that this model is incapable of describing the correct level splitting due to fine- and hyperfine splitting observed experimentally \cite{Thewes2015}. For this reason, the simple hydrogenic theory has been expanded taking into account the complex valence band structure and the cubic symmetry $O_{\rm h}$ of $\textrm{Cu}_2\textrm{O}$ in a quantitative theoretical framework \cite{Lipari1970,Lipari1973a,Lipari1973b,Uihlein1981,Suzuki1974,Schoene2016}. 

The addition of external electric and magnetic fields reduces the symmetry of the system, therefore leading to level structures possessing numerous complex splitting of excitonic absorption lines \cite{Agekyan1977}. The analysis of excitonic absorption spectra in both electric and magnetic field strengths has been a long-standing subject from the theoretical as well as experimental point of view \cite{Altarelli1972,Cho1974,Semina2016a,Semina2016b}. Due to specific material parameters, excitonic properties such as Bohr radius and electric/magnetic field strength units provide the possibility to access exotic regimes more easily compared to standard atomic systems. For instance, recent high-resolution spectroscopy and intensive theoretical calculations of excitons in $\textrm{Cu}_2\textrm{O}$ have provided a fundamental understanding of complex excitonic absorption spectra in external magnetic fields for applied field strengths of up to $7\, \rm T$ and excitonic states with principal quantum numbers $n \le 7$ \cite{Schweiner2016,Schweiner2016b}. 

In atomic physics, an exotic species of highly excited Rydberg states in crossed electric and magnetic fields are the so-called giant-dipole atoms. This particular atomic species has been predicted theoretically \cite{Baye1992, Dzyaloshinskii1992,Schmelcher1993a,Dippel1994, Schmelcher1998,Schmelcher2001} and explored experimentally in the early 1990's \cite{Fauth1987,Raithel1993}. When the center-of-mass and relative motion of the field-dressed species are treated correctly, the total momentum of the system is not a conserved quantity and an exact separation of the atomic degrees of freedom is impossible \cite{Schmelcher1993a}. The pseudomomentum is, however, a conserved quantity and for neutral systems one can carry out a pseudoseparation of the center-of-mass  and relative motion. It has been shown that the effect of the center-of-mass degrees of freedom on the internal motion is an effective potential that gives rise to an outer well for certain values of the pseudomomentum and applied field strengths. 

This leads to delocalized states, the so-called giant-dipole states. In contrast to the usual Rydberg states, giant-dipole states are of decentered character with an electron-ionic core separation up to several micrometers, leading to huge permanent electric dipole moments in the range of hundreds of thousand Debye. Applications to matter-antimatter atoms have predicted bound state lifetimes of many years, and recent studies have indicated the existence of diatomic ultra-long ranged giant-dipole molecules \cite{Kurz2012}. 

However, the concept of giant-dipole atoms is not restricted to real atomic systems as it can, in principle, be applied to neutral quasi-particle systems such as excitons as well. For instance, Schmelcher analyzed excitons with non-vanishing pseudomomentum in an external magnetic field within an effective hydrogenic model \cite{Schmelcher1993ex}. As the simple hydrogen-like approach has turned out to be insufficient to describe both the field-free as well as the field-dressed excitonic species, it is obvious that a more complex theoretical approach is required to derive a sufficient description of possible excitonic giant-dipole states. Therefore, in the present work we expand the concept of atomic giant-dipole states to realistic semiconductor environments. Starting from the exact field-dressed Hamiltonian we derive the theoretical foundation of excitons in crossed fields. We then consider $\textrm{Cu}_2\textrm{O}$ and calculate the specific properties of giant-dipole potentials in this material.

This paper is organized as follows. In Sec.\ \ref{gd_Hamiltonian} we present the Hamiltonian of excitons in crossed electric and magnetic fields. Performing a gauge-independent pseudoseparation of the center-of-mass and relative motion we derive an effective single-particle description of the field-dressed excitonic system. As a result we obtain a spatially dependent electron-hole interaction potential. Furthermore, we show the possibility of Abelian and non-Abelian gauge field description of field-dressed excitonic systems. In Sec.\ \ref{gd_potential} we derive the potential energy surfaces of the excitonic giant-dipole system for various electric and magnetic field strengths and orientations. We obtain several potential surfaces providing possible electron-hole separation up to several micrometers. We show that by varying both the electric and magnetic field strengths one can easily change the topological properties of the potential surfaces. Finally, we give a short summary and outlook in Sec. \ref{conclusion}.
\section{The excitonic Hamiltonian in external electric and magnetic fields \label{gd_Hamiltonian}}
The Wannier excitons in $\rm Cu_2O$ which are analyzed throughout this work are formed by an electron in the lowest $\Gamma^{+}_{6}$-conduction band and a positively charged hole in the uppermost $\Gamma^{+}_{5}$-valence band. Here, the latter is triply degenerate. The energy gap between the two bands is given as $E_g=2.17208\, \rm eV$ \cite{Scheel2014}. As the $\Gamma^{+}_{6}$-band is almost parabolic in the vicinity of the $\Gamma$-point, the kinetic energy 
\begin{eqnarray}
H_e(\bfmath{p}_e)=\frac{\bfmath{p}^{2}_{e}}{2m_e}
\end{eqnarray}
of the electron is determined by an isotropic effective mass $m_e=0.985m_0$ which is almost identical to the free electron mass $m_0$. 

\begin{figure}[h]
\centering
\begin{minipage}{0.475\textwidth} 
\includegraphics[width=\textwidth]{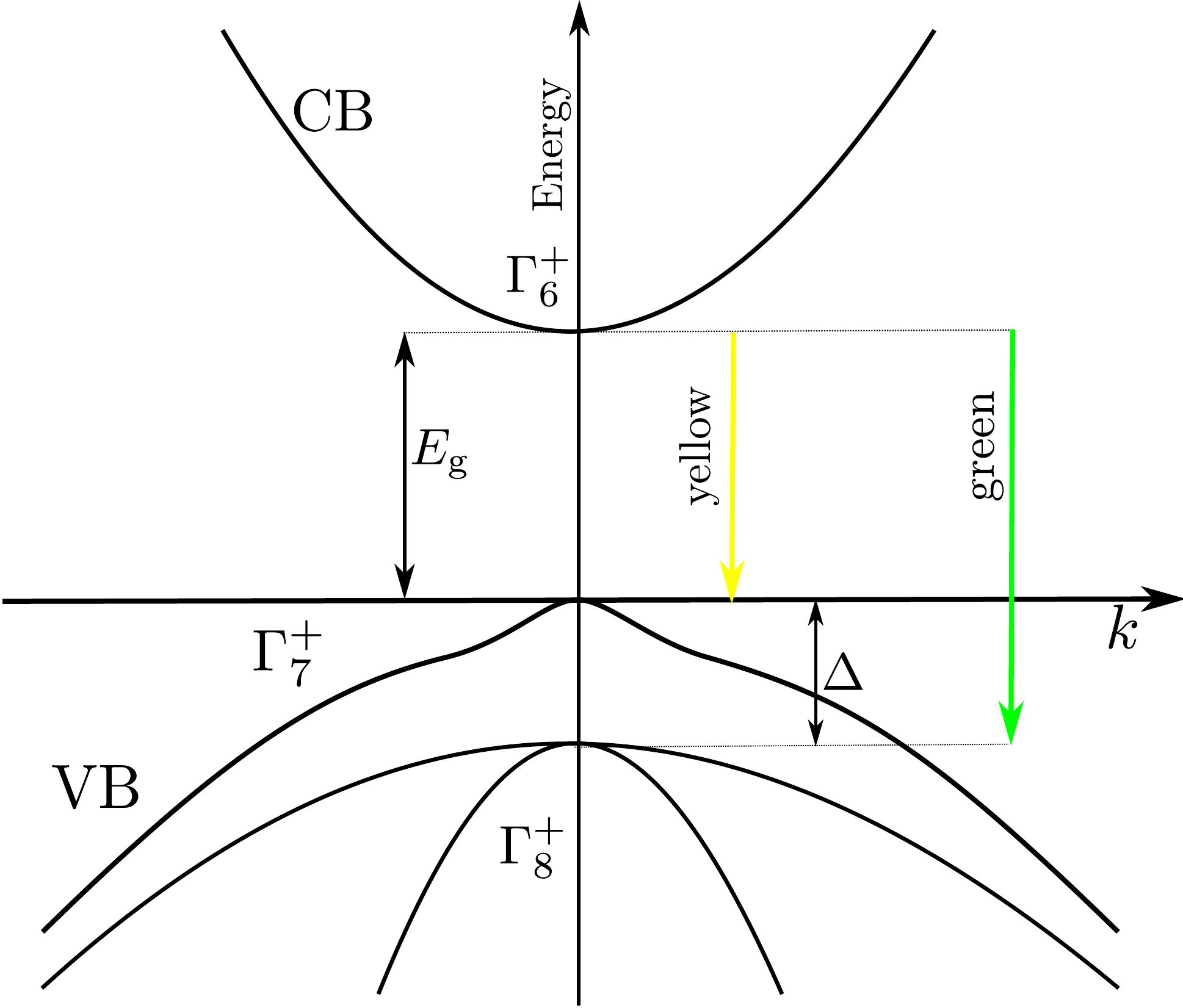}  
\end{minipage}
\caption{Schematic band structure in $\textrm{Cu}_2\rm O$. Transitions between the conduction band (CB) and valence bands (VB) lead to two excitonic series denoted as yellow and green.}
\label{plot1a}
\end{figure}

In contrast to the conduction band, the three uppermost valence bands are deformed due to interband interactions and non-spherical symmetry properties of the solid. These properties can be represented by an effective $I=1$ quasi-spin representation in the hole degrees of freedom \cite{Luttinger1956}. Thus, the kinetic energy Hamiltonian $H_h(\bfmath{p}_h)$ of a hole in the case of three coupled valence bands is given by a more complex expression determined by the three Luttinger parameters $\gamma_i,\ i=1,2,3$ \cite{Luttinger1954,Suzuki1974}
\begin{eqnarray}
H_{h}(\bfmath{p}_h)&=&\frac{\bfmath{p}^{2}_{h}}{2m_0}(\gamma_1+4\gamma_2)-\frac{3\gamma_2}{m_0}(p^{2}_{h,x}I^{2}_{x}+{\rm c.p.})\nonumber\\
&\ &-\frac{6\gamma_3}{m_0}[\{p_{x,h}p_{y,h}\}\{I_xI_y\}+{\rm c.p.}].\label{Ham_hole}
\end{eqnarray}
The mapping $\{a,b\}=\left( ab+ba \right)/2$ is the symmetric product and c.p.\ denotes cyclic permutations \cite{Suzuki1974}. It can be used to define the elements of a symmetric and trace-free Cartesian spin tensor with elements $I_{ij}$
\begin{eqnarray}
I_{ij}=3\{ I_i,I_j \}-2\delta_{ij}1_I,\ \ \ i=x,y,z. \label{I_cart}
\end{eqnarray}
The operator $1_I$ denotes the unity operator of the $I=1$ pseudo-spin representation. Including the hole spin $\bfmath{S}_h$ with $S_h=1/2$, each of the three $\Gamma^{+}_{5}$-valence bands becomes doubly degenerate. However, the quasi-spin $\bfmath{I}$ not only changes the kinetic energy term of the hole but also effectively couples to the total effective hole spin $\bfmath{J}=\bfmath{I}+\bfmath{S}_h$. Because of the spin-orbit coupling, the degenerate valence bands split into one single higher-lying doubly degenerate $\Gamma^{+}_{7}$ and two doubly degenerate lower-lying $\Gamma^{+}_{8}$-valence bands separated by an amount of $\Delta=133.8\, \rm meV$ (see Fig.\ \ref{plot1a}). Optical transitions between the conduction band and the two valence bands provide two distinct optical series, namely the yellow ($J=1/2$) and green series $(J=3/2)$, respectively (see Fig.\ \ref{plot1a}). 

Throughout this paper, the ionization threshold of the yellow series is chosen to be the zero point of the energy scale.
Furthermore, if not stated otherwise, we use excitonic Hartree units, i.e.\ $e=\hbar=m_0/\gamma^{'}_{1}=1/4\pi\epsilon_0 \varepsilon=1$. Here, $\varepsilon=7.5$ denotes the static dielectric constant of the bulk material and $\gamma^{'}_{1} \equiv m_0/m_e+\gamma_1$. In this particular unit system, the energies are measured in units of the excitonic Hartree energy, $\mathcal{H}_{\rm ex}=174\, \rm meV$, while the distances are measured in units of the corresponding excitonic Bohr radius, $a_{\rm ex}=\gamma^{'}_{1}\varepsilon a_0$, where $a_0$ is the atomic Bohr radius. In table (\ref{table}) a detailed list of the physical quantities considered in the present work is presented.\\
\begin{table}[h]
\centering
\begin{tabular}{ c c c }
  \hline			
Hartree energy & $\mathcal{H}_{\rm ex}$ &$174\, \rm meV$ \\
(excitonic) Bohr radius & $a_{\rm ex}$ & $1.1\, \rm nm$\\
  magnetic field strength & $B_{\rm ex}$ & $542.5\, \rm T$ \\
  electric field strength & $E_{\rm ex}$ & $1.583\, \rm MV/cm$ \\
  momentum & $P_{\rm ex}$ & $4.8 \times 10^{-2}\hbar/a_0$ \\
  dipole moment & $d_{\rm ex}$ & $52.96\, \rm D^{1}$\\
  gap energy & $E_g$ & $2.17208\, \rm eV$\\
  spin-orbit coupling &$\Delta$ & $133.8\, \rm meV$\\
Bohr magneton &$\mu_B$& $57.88\, \mu \rm eV/T$\\
Luttinger parameters & $\gamma_1,\gamma_2,\gamma_3$ & $1.76,0.82,0.54$\\
& $ \gamma^{'}_{1},\kappa$ & $2.78,-0.5$  \\ 
  \hline  
\end{tabular}
\caption{Excitonic Hartree energy $\mathcal{H}_{\rm ex}$, Bohr radius $a_{\rm ex}$, external field strengths $(B_{\rm ex},E_{\rm ex})$, momentum $K_{\rm ex}$ and electric dipole moment $d_{\rm ex}$ expressed in commonly used units. In addition, the spin-orbit and magnetic coupling $(\Delta,\mu_B)$ is presented as well as the Luttinger parameters used throughout this work.}
\label{table}
\end{table}
In case an external magnetic field is applied, the canonical momenta of electron and hole are replaced by their kinetic momenta $\bfmath{p}_{e/h} \rightarrow \bfmath{p}_{e/h}\pm A(\bfmath{r}_{e/h})$, where $\bfmath{A}(\bfmath{r})$ is the vector potential, and the magnetic field is given by $\bfmath{B}(\bfmath{r})=\bfmath{\nabla}\times \bfmath{A}(\bfmath{r})$. Obviously, the vector potential is not uniquely defined but can be gauged using the gradient of a scalar field $\Lambda(\bfmath{r})$: $\bfmath{A}'(\bfmath{r})=\bfmath{A}(\bfmath{r})+\grad \Lambda(\bfmath{r})$. For a homogeneous magnetic field, the vector potential in an arbitrary gauge can be written as $\bfmath{A}(\bfmath{r})=\bfmath{A}_{\rm sym}(\bfmath{r})+\grad \Lambda(\bfmath{r})$ with $\bfmath{A}_{\rm sym}(\bfmath{r})=\frac{1}{2}\bfmath{B}\times \bfmath{r}$ denoting the symmetric gauge. In case that a homogeneous external electric field is applied as well, the Stark terms $\mp \bfmath{E} \cdot \bfmath{r}_{e/h}$ of the electron and hole have to be considered in addition. For this reason, the Hamiltonian of excitons in homogeneous external electric and magnetic fields is given by\let\thefootnote\relax\footnote{$^{1}1\, \textrm{D}=0.393ea_0$, i.e. $1\, \textrm{D}/e=20.8\,\rm pm$.}
\begin{eqnarray}
H_{\rm ex}&=&\frac{1}{2m_e}(\bfmath{p}_{e}+\bfmath{A}_{\rm sym}(\bfmath{r}_e)+\bfmath{\nabla}_e \Lambda_e)^2-\frac{1}{|\bfmath{r}_e-\bfmath{r}_h|}\nonumber\\
&\ &+H_h(\bfmath{p}_{h}-\bfmath{A}_{\rm sym}(\bfmath{r}_h)-\bfmath{\nabla}_h \Lambda_h)+\bfmath{E}\cdot(\bfmath{r}_e-\bfmath{r}_h)\nonumber\\
&\ &+H_{\rm so}+H_B \label{Ham_exc}
\end{eqnarray}
with $m_e \rightarrow m_e\gamma^{'}_{1}/m_0$, $\grad_i \Lambda_i \equiv \grad_{\bfmath{r}_i}\Lambda(\bfmath{r}_i)$ and
\begin{eqnarray}
H_{\rm so}&=&\frac{2}{3}\bar{\Delta} (1+\bfmath{I}\cdot\bfmath{S}_h),\\ 
H_B&=&\bar{\mu}_B [(3\kappa+\frac{g_s}{2})\bfmath{I}\cdot\bfmath{B}-g_s\bfmath{S}_{h}\cdot\bfmath{B}],\\ \bar{\Delta}&\equiv&\frac{\Delta}{\mathcal{H}_{\rm ex}},\ \ 
\bar{\mu}_{B}\equiv\frac{\mu_B}{\mathcal{H}_{\rm ex}}. 
\end{eqnarray}
The term $H_{\rm so}$ denotes the spin-orbit coupling of the hole spin $\bfmath{S}_{h}$ with $\bfmath{I}$, while $H_B$ includes the coupling of the hole spins to the external magnetic field. As we do not include any kind of electronic spin-orbit coupling or spin-spin interaction, the electron spin $\bfmath{S}_{e}$ is not considered throughout this work. The quantity $\bar{\mu}_B\approx0.18/B$ denotes the (scaled) Bohr magneton and $g_s \approx 2$ is the $g$-factor of the hole spin. The value of the Luttinger parameter $\kappa$ has been determined recently to be $\kappa = -0.5 \pm 0.1$ via high-resolution spectroscopy of magnetoexcitons in $\textrm{Cu}_2\textrm{O}$ \cite{Schweiner2016}.

Next, we introduce the center-of-mass vector $\bfmath{R}$ and the relative vector $\bfmath{r}=\bfmath{r}_e-\bfmath{r}_h$. As it has been discussed in previous works, the excitonic Hamiltonian $H_{\rm ex}$ possesses a constant of motion, the so-called pseudomomentum $\hat{\bfmath{K}}$, which is given by \cite{Avron1978,Herold1981,Johnson1983}
\begin{eqnarray}
\hat{\bfmath{K}}=\bfmath{P}-\frac{1}{2}\bfmath{B}\times \bfmath{r}+\grad_{\bfmath{R}}(\Lambda_h-\Lambda_e).\label{K_pseudo}
\end{eqnarray}
In static magnetic fields, the components of the pseudomomentum commute with the excitonic Hamiltonian (\ref{Ham_exc}). In addition, for neutral systems, the components of $\hat{\bfmath{K}}$ commute with one another. Therefore, the eigenfunctions of the corresponding excitonic Schr\"odinger equation can be chosen as simultaneous eigenfunctions of the pseudomomentum \cite{Dippel1994}. As it has been shown by Dippel {\it et al.} \cite{Dippel1994}, in this case the Hamiltonian can be transformed via a unitary transformation into a single-particle Hamiltonian
\begin{eqnarray}
H_{\rm ex}&=&\frac{1}{2m_e}(\bfmath{p}+\bfmath{A}_{\rm sym}(\bfmath{r})+\frac{m_e}{M}\bfmath{K}+\grad f(\bfmath{r}))^2-\frac{1}{r}\nonumber\\
&\ &+H_h(\bfmath{p}-\bfmath{A}_{\rm sym}(\bfmath{r})-\frac{m_h}{M}\bfmath{K}+\grad f(\bfmath{r}))+\bfmath{E}\cdot\bfmath{r}\nonumber 
\\&\ &+H_{\rm so}+H_B \label{Ham_exc_rel}
\end{eqnarray}
where $f(\bfmath{r})$ is a function of the relative coordinate only whose gradient simply reflects the gauge freedom of the relative motion's vector potential. The quantity $M=m_e+m_h,\ m_h \equiv m_0/\gamma_1$ is the total excitonic mass. Furthermore, the vector $\bfmath{K} \in \mathbb{R}^3$ denotes the vector of eigenvalues of the pseudomomentum components.

The single-particle Hamiltonian given in Eq.\ (\ref{Ham_exc_rel}) is not the final expression of the effective Hamiltonian as it still mixes kinetic and potential terms. In order to decouple the kinetic and potential energy terms, we introduce the kinetic momentum $\bfmath{\pi}$ with
\begin{eqnarray}
&\ &\hspace{-0.75cm}\pi_i=1_Ip_i+1_I\partial_i f \nonumber\\
&\hspace{-0.25cm}-&\sum_{k}[(\frac{m_h}{M}1_I\delta_{ki}-\Omega_{ki})K_k+(1_I\delta_{ki}-2\Omega_{ki})A^{(k)}_{\rm{sym}}], \label{pi}
\end{eqnarray}
where the matrix elements $\Omega_{ij}$ are considered to be spin matrices and $\partial_i \equiv (\grad)_i$. For a system of two equally charged particles we obtain $\Omega_{ij}=1_I\delta_{ki}m_h/M$. Therefore, the $\bfmath{K}$-dependent term vanishes and we exactly reproduce the result of Ref.\ \cite{Dippel1994}. As the operators $I_{ij}$ form a closed subset with respect to the symmetric product $\{ a,b \}$, the matrix elements $\Omega_{ij}$ can be expanded in the following way \cite{Schweiner2016b}: 
\begin{eqnarray}
\Omega_{jj}=C_11_I+\frac{C_2}{3}I_{jj},\ \ \ \Omega_{jk}=\frac{C_3}{3}I_{jk},\ \ j \not= k.
\end{eqnarray}
The coefficients $C_i \in \mathbb{R}$ are functions of the electron mass $m_e$ and the Luttinger parameters $\gamma_i$ and have been derived in Ref.\ \cite{Schweiner2016b}. Inserting Eq.\ (\ref{pi}) into Eq.\ (\ref{Ham_exc_rel}), the excitonic Hamiltonian $H_{\rm ex}$ takes the form
\begin{eqnarray}
&\ &\hspace{-0.75cm}H_{\rm ex}=\frac{\bfmath{\pi}^2}{2m_e}+H_{h}(\bfmath{\pi})+V(\bfmath{r})+\frac{2}{3}\bar{\Delta}(1+\bfmath{I}\cdot\bfmath{S}_h) \nonumber\\
&\ &+\bar{\mu}_B [(3\kappa+\frac{g_s}{2})\bfmath{I}\cdot\bfmath{B}-g_s\bfmath{S}_{h}\cdot\bfmath{B}]\label{Ham_exc_final},
\end{eqnarray}
with
\begin{eqnarray}
\hspace{-0.5cm}V(\bfmath{r})&=&\left(\Omega_1 \tilde{K}^2+\bfmath{E}\cdot\bfmath{r}-\frac{1}{r}\right) 1_I-\Omega_2\sum_i\tilde{K}^{2}_{i}I_{ii}\nonumber\\ &\ &-\frac{2}{3}\Omega_3\sum_{ij,j<i}\tilde{K}_{i}\tilde{K}_{j}I_{ij},\ \tilde{\bfmath{K}}=\bfmath{K}+\bfmath{B}\times \bfmath{r}.\label{gd_V}
\end{eqnarray}
The coefficients $\Omega_i,\ i=1,2,3$ are given in the Appendix.

In the final excitonic Hamiltonian, Eq.\ (\ref{Ham_exc_final}), the only gauge-dependent terms are the expressions depending on the kinetic momentum $\bfmath{\pi}$. Obviously, these terms can be associated with the kinetic energy of the internal motion. Within this expression, we define the components of the effective vector potential $\bfmath{A}_{\rm eff}$ as
\begin{eqnarray}
A^{(i)}_{\rm eff}(\bfmath{r})=\sum_{k}(1_I\delta_{ik}-2\Omega_{ik})A^{(k)}_{\rm sym}(\bfmath{r})-1_I\partial_{i} f
\label{Aeff}.
\end{eqnarray}
According to the Berry connection \cite{Kiffner2013,Kiffner2013b}, this vector potential is related to an effective magnetic field $\bfmath{B}_{\rm eff}$ given by
\begin{eqnarray}
&\ &\hspace{-1em}(B^{(i)}_{\rm eff})_{nm}=\frac{1}{2}\epsilon_{ikl}F^{(kl)}_{nm},\nonumber \\
&\ &\hspace{-1em}F^{(kl)}_{nm}=\partial_k (A^{(l)}_{\rm eff})_{nm} - \partial_l (A^{(k)}_{\rm eff})_{nm} -i [A^{(k)}_{\rm eff},A^{(l)}_{\rm eff}]_{nm}.\label{Beff}
\end{eqnarray}
As the matrix elements $\Omega_{ij}$ are linear combinations of the spin matrices $I_{ij}$, the components $A^{(i)}_{\rm eff}$ of the vector potential do no commute in general. In fact, this property reflects the non-Abelian character of the effective gauge potential, Eq.\ (\ref{Aeff}). In contrast to the gauge dependent terms of the kinetic energy operator, the term $V(\bfmath{r})$, according to Eq.\ (\ref{gd_V}), is gauge independent. Therefore, it represents an effective single-particle potential for the internal motion. 

Note that, so far, we have made no approximation within the analysis of the field-dressed excitonic species. For this reason, the excitonic Hamiltonian, Eq.\ (\ref{Ham_exc_final}), provides the full dynamics of electrically and magnetically field-dressed excitons.
\section{Giant-Dipole Potential Surfaces \label{gd_potential}}
In this section, we analyze the properties of the potential $V(\bfmath{r})$ of the internal motion in more detail. Due to the spin $I=1$ degree of freedom, the potential can be expressed as a $3 \times 3$ matrix where the matrix elements are functions of the external field parameters $\bfmath{B}$ and $\bfmath{E}$ and the spatial coordinate $\bfmath{r}$, respectively (see Appendix). As the potential matrix is Hermitian, its diagonalization provides three real eigenvalues $\varepsilon_i,\ i=1,2,3$. Taking the external fields as parameters, the spatially dependent eigenenergies define potential energy surfaces $V_{i}(\bfmath{r}) \equiv \varepsilon_i(\bfmath{r};\bfmath{B},\bfmath{E},\bfmath{K})$. Obviously, the potential surfaces can be obtained analytically for arbitrary field strengths and configurations. Considering the hole spin $\bfmath{S}_h$ in addition, one observes that the potential $V(\bfmath{r})$ does not couple any hole spin degrees of freedom, which means the eigenvalues $\varepsilon_i$ are doubly degenerate. 

Throughout this work, we consider the electric field to be oriented along the $[001]$ direction. Furthermore, we always assume the fields to be oriented perpendicular to one another, i.e.\ $\bfmath{B}\perp \bfmath{E}$. 

\subsection{Magnetic field in [100] direction}
\subsubsection{Perturbative analysis}
In the case of the magnetic field oriented along the $[100]$ direction, the expressions for the potential energy surfaces are more compact and are given by
\begin{eqnarray}
V_1(\bfmath{r})&=&\left( \Omega_1- \Omega_2 \right)\tilde{K}^2 +Ez-\frac{1}{ r},\nonumber\\
V_{2,3}(\bfmath{r})&=&\left( \Omega_1- \Omega_2 \right)\tilde{K}^2 +Ez-\frac{1}{ r}+\frac{3}{2}\Omega_{2}\left( \tilde{K}^{2}_{2}+\tilde{K}^{2}_{3} \right)\nonumber \\ 
&\ & \pm \frac{1}{2}\sqrt{9\Omega^{2}_{2}\left( \tilde{K}^{2}_{2}-\tilde{K}^{2}_{3} \right)^2 + 4\Omega^{2}_{3}\tilde{K}^{2}_{2}\tilde{K}^{2}_{3}}. \label{Vgd_analytic}
\end{eqnarray}
The potential $V_1(\bfmath{r})$ can be identified to be the potential term discussed by Dippel {\it et.\ al} for a system of two charged particles in crossed electric and magnetic fields \cite{Dippel1994}. For such a system, the coefficient $ \Omega_1- \Omega_2$ is replaced by $1/2\mathcal{M}$ where $\mathcal{M}$ denotes the total mass of the atomic system. Because
\begin{eqnarray}
\tilde{K}^2=B^2(z^2+y^2)+2(\bfmath{K} \times \bfmath{B})\bfmath{r}+K^2 \label{Ktilde},
\end{eqnarray}
we see that the topology of this particular potential surface is determined by the diamagnetic-like term $\sim B^2$ as well as the external electric field $\bfmath{E}$ and the so-called motional electric field $2(\bfmath{K} \times \bfmath{B})(\Omega_1-\Omega_2)$, respectively. In Ref.\ \cite{Dippel1994}, it has been shown that the Stark term $\bfmath{E}\cdot\bfmath{r}$ can be included into $\tilde{\bfmath{K}}$, which leads to a shift of the pseudomomentum, $\bfmath{K}\rightarrow\bfmath{K}-M\bfmath{v}_D$. Here, the additional term includes the drift velocity $\bfmath{v}_D=\bfmath{E}\times \bfmath{B}/B^2$ of the charged particles in crossed fields. For the excitonic system under consideration, such a replacement is not possible as the electric field only appears in the diagonal elements of the interaction potential $V(\bfmath{r})$ whereas the magnetic terms appear in all matrix elements (see Eq.\ (\ref{gd_V})). For this reason, we threat the electric field $\bfmath{E}$ and the pseudomomentum $\bfmath{K}$ as independent parameters. 
\begin{figure}[h]
\begin{minipage}{0.45\textwidth} 
\includegraphics[width=\textwidth]{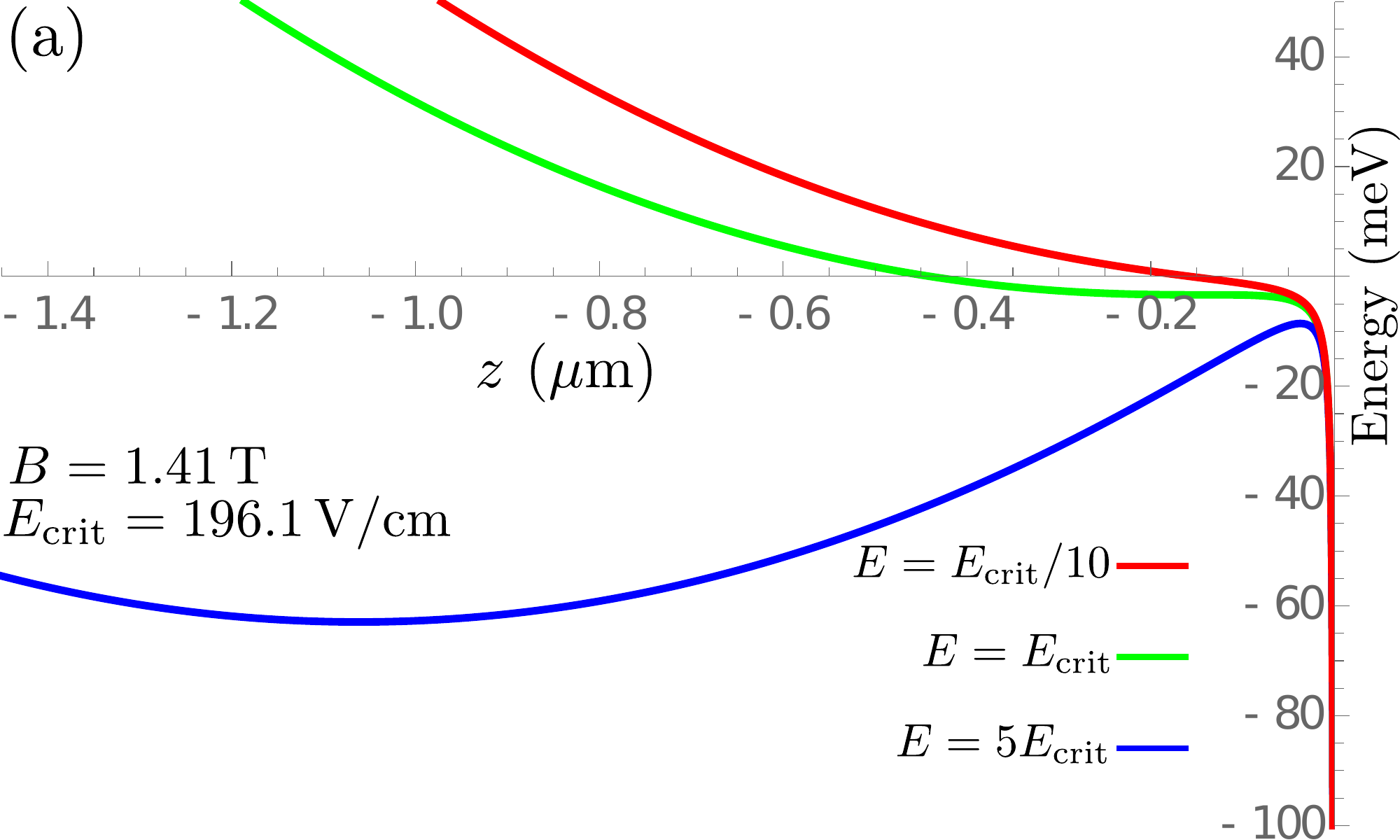}  
\end{minipage}
\begin{minipage}{0.45\textwidth} 
\includegraphics[width=\textwidth]{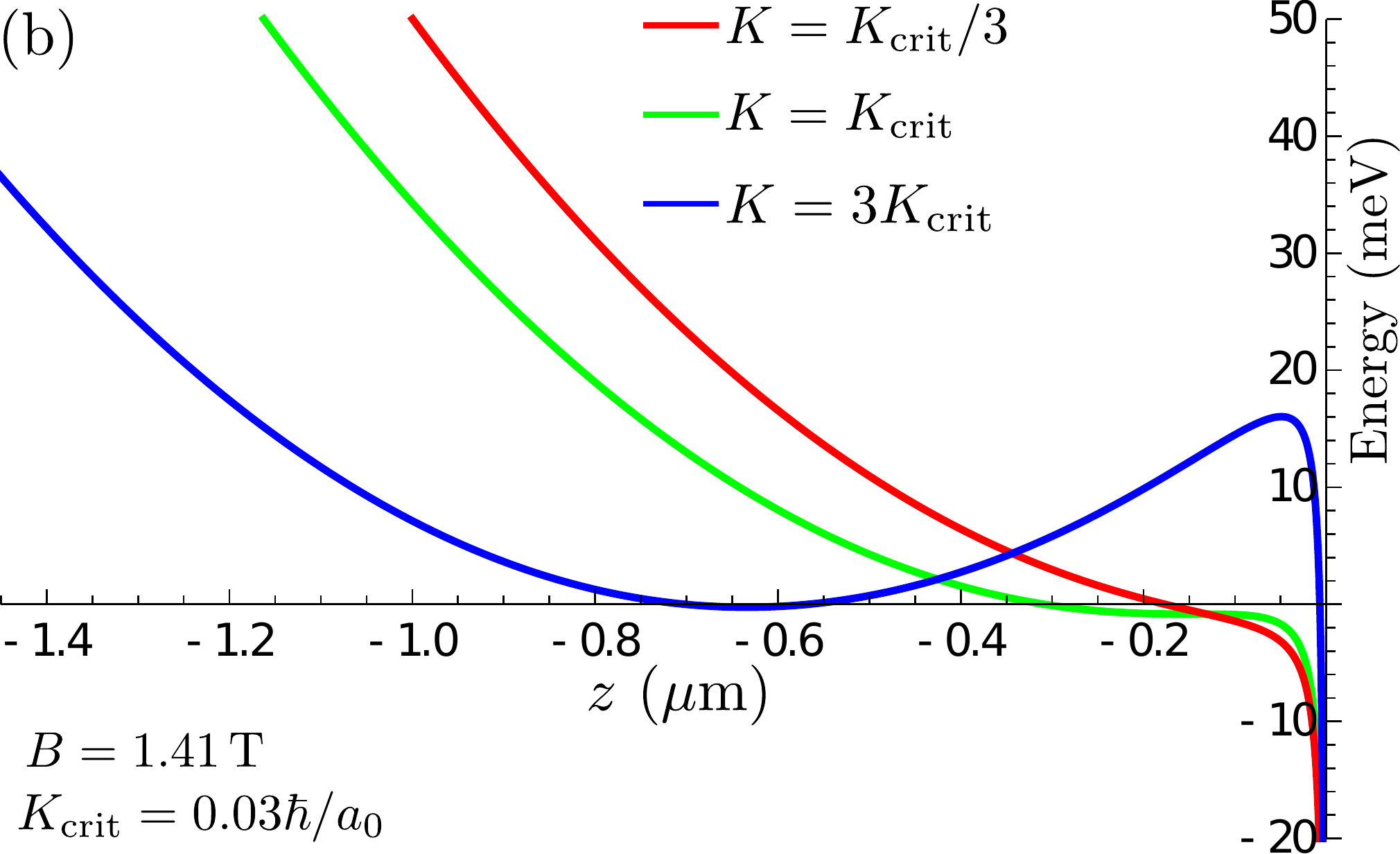}  
\end{minipage}
\caption{(a) Potential cuts of the potential surface $V_{1}(\bfmath{r})$, $y=z=0$ for various electric field strength $E$ and $B=1.41\, \rm T$, which gives $E_{\rm crit}=196.3\, \rm V/cm$. The potential curve for $E=5E_{\rm crit}$ possesses a minimum around $z^{(1)}_{\rm min} \approx -1.05\, \mu \rm m$ and a lowering of around $52.5 \rm meV$ with respect to the saddle point.  For $E=E_{\rm crit}$ (green solid line) a single plateau is present instead of a potential minimum. In case of $E=E_{\rm crit}/10$ (red solid curve) the plateau has vanished. (b) Potential curves for $E=0$ but finite $K$. For $K=3K_{\rm crit}$ (blue solid line) a potential minimum $z^{(1)}_{\rm min} \approx 0.65\, \mu \rm m$ is present. With decreasing $K$ the minimum turns into a plateau ($K=K_{\rm crit}$, green curve) until the potential curve is monotonically decreasing ($K=K_{\rm crit}/3$, red curve). For $B=1.41\, \rm T$ one gets $K_{\rm crit}=0.03\hbar/a_0$.}
\label{plot1}
\end{figure}

First, we set $\bfmath{K}=0$, in this case the Stark term related to the external electric field alone compensates the quadratic growth of the magnetic term for sufficiently small spatial separations $\bfmath{r}$. For $|\bfmath{r}| \rightarrow 0$, the Coulomb singularity  becomes the dominant part in the interaction potential. Due to the interplay between electric and magnetic fields, we expect potential surfaces which provide outer local minima. From the condition for a potential minimum $\bfmath{\nabla}V_{i}(\bfmath{r})=0$ we obtain $y_{\rm min}=x_{\rm min}=0$ for all three surfaces. Because $V_{1}(z)=V_{3}(z)$ for $y=x=0$, we are left with two cubic equations for the $z$-coordinate, namely
\begin{eqnarray}
2B^2(\Omega_1+\frac{\Omega_2}{2}(1 \pm 3))z^{3}+Ez^{2}-1=0.
\end{eqnarray}
It turns out that, beyond a specific critical electric field strength $E_{\rm crit}= 3\sqrt[3]{B^{4}(\Omega_1-\Omega_2)^2}$, the three-dimensional potential surface $V_{1}(\bfmath{r})$ possesses both a minimum $z^{(1)}_{\rm min}$ and a saddle point $z^{(1)}_{\rm s}$ with
\begin{eqnarray}
z^{(1)}_{\rm min}&=&\frac{E}{6(\Omega_1-\Omega_2)B^2}[2\cos(\frac{\theta+2\pi}{3})-1]\nonumber, \\
z^{(1)}_{\rm s}&=&\frac{E}{6(\Omega_1-\Omega_2)B^2}[2\cos(\frac{\theta+4\pi}{3})-1]\label{positions}
\end{eqnarray}
with $\cos(\theta)=54(\Omega_1-\Omega_2)^2B^4/E^3-1$ (see Ref.\ \cite{Schmelcher1998}). In contrast, the potential surface $V_{2}(\bfmath{r})$ only possesses a saddle point and no local minima.

In Fig.\ (\ref{plot1}a), we present cuts of the three potential curves $V_{1}(0,0,z)\equiv V_{1}(z)$ and different electric field strengths. The magnetic field strength is set to $B=1.41\, \rm T$, which gives a critical electric field strength of $E_{\rm crit} = 196.3\, \rm V/cm$. Because of $y=0$, we have $V_{1}(z)=V_{3}(z)$. In Fig.\ (\ref{plot1}a), the blue solid curve represents the potential for an applied electric field strength of $E=5E_{\rm crit}$. Clearly, $V_1(z)$ possesses a pronounced potential minimum at $z^{(1)}_{\rm min} \approx -1.05\, \mu \rm m$ and a lowering of $52.5\, \rm meV$ with respect to the saddle point. In contrast, for an applied field strength of $E=E_{\rm crit}$ (green solid line), the potential minimum of the curve $V_1(z)$ has vanished and only a plateau at the saddle point position $z^{(1)}_{\rm s} \approx 174\, \rm nm$ is present. Finally, the red solid line represents the situation for even weaker field strength. In particular, this curves shows $V_{1}(z)$ for $E=E_{\rm crit}/10$. We clearly see that the plateau has vanished and the curve is monotonically increasing for $z \rightarrow -\infty$.
\begin{figure}[h]
\centering
\begin{minipage}{0.45\textwidth} 
\includegraphics[width=\textwidth]{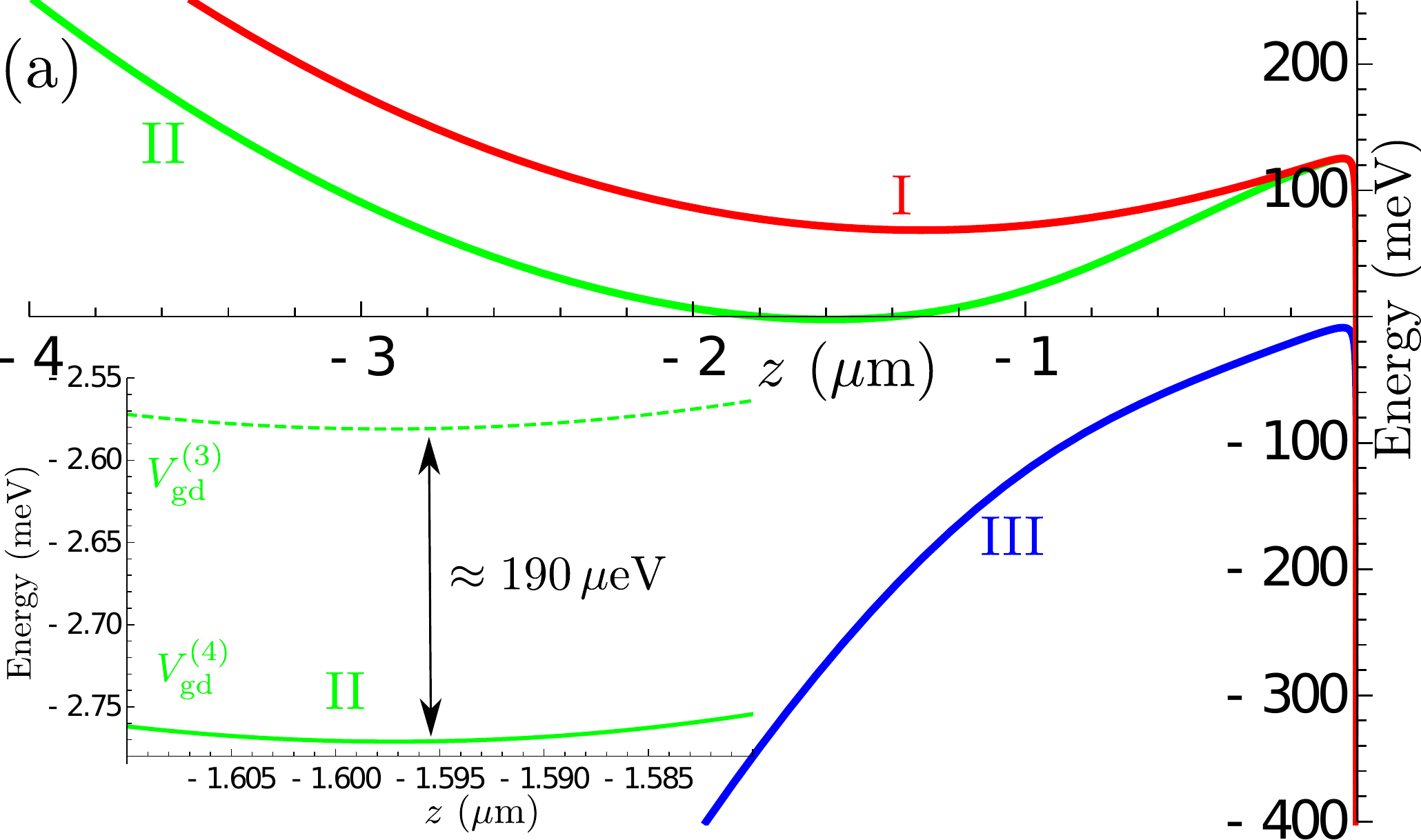}  
\end{minipage}
\begin{minipage}{0.45\textwidth} 
\includegraphics[width=\textwidth]{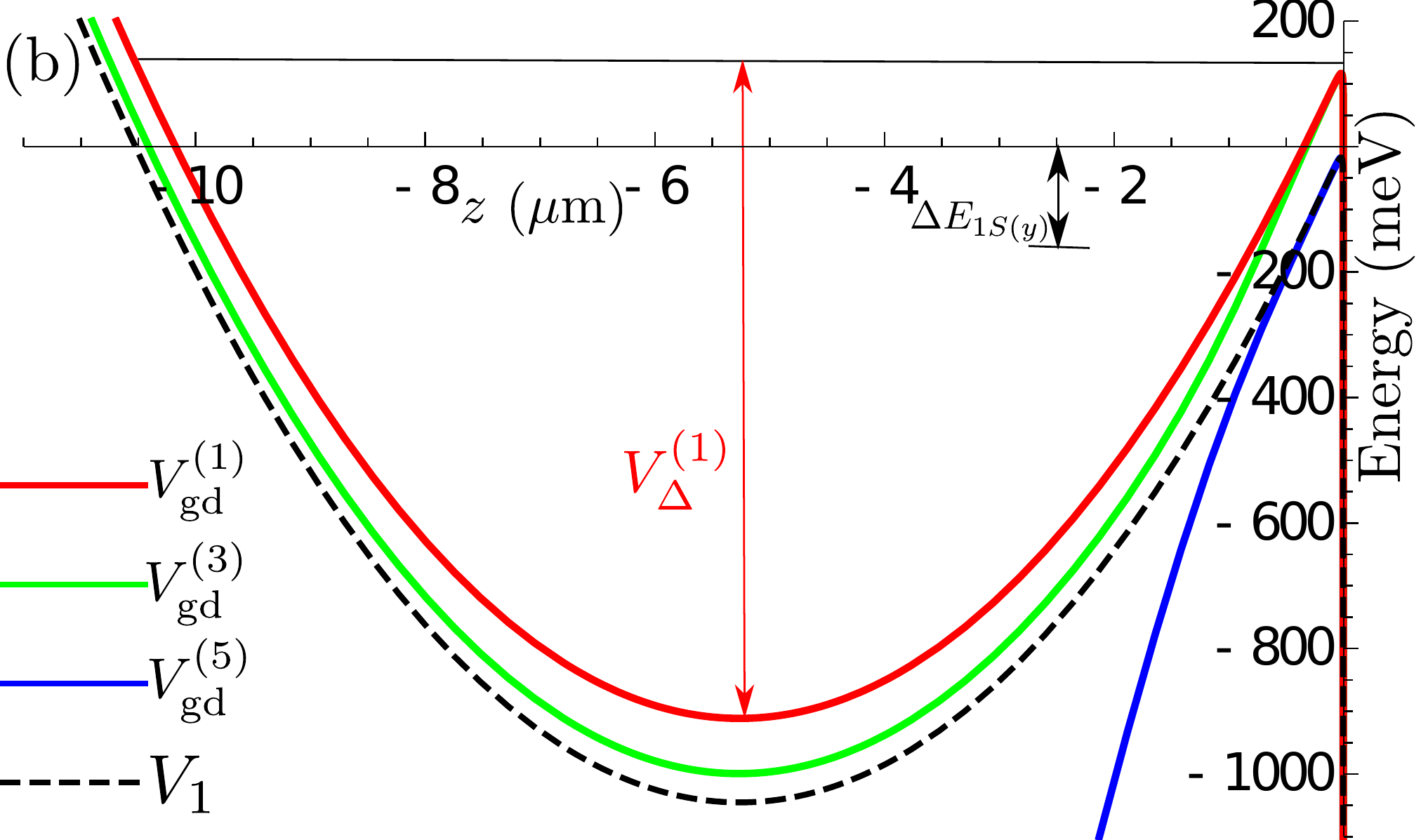} 
\end{minipage}
\caption{(a) Giant-dipole potential curves for applied field strengths $B=1.41\, \rm T$, $E=1\, \rm kV/cm$ plotted along the $z$-axis ($y=x=0$). Due to the small spacing of approximately $190\, \mu \rm eV $ only three of six potential curves $V^{(i)}_{\rm gd}$ denoted with I,II,III are visible (see inset). (b) Potential curves for $B=1.41\, \rm T$, $E=4\, \rm kV/cm$ (solid curves). The spacing $V^{(1)}_{\rm \Delta}$ is indicated as well as the binding energy $\Delta E_{1S(y)}$ of the $1S$-yellow ground state exciton ($\Delta E_{1S(y)}=150\, \rm meV$). In addition, the analytic potential curve $V_{1}(z)$ is shown in comparison (black dashed curve).}
\label{plot2}
\end{figure}

Next we analyze the system for zero electric field ($E=0$) but finite pseudomomentum $\bfmath{K}=(0,K,0)^T$. According to Eq.\ (\ref{Ktilde}), this situation is analogous to the previous calculations where we considered a finite Stark term. We obtain a critical pseudomomentum $K_{\rm crit}=\frac{3}{2}\sqrt[3]{B/(\Omega_1-\Omega_2)}$ for the existence of a potential minimum. The position of the minimum is easily obtained by the replacement $E/(2(\Omega_1-\Omega_2)B^2) \rightarrow K/B$ in Eq.\ (\ref{positions}). In Fig.\ (\ref{plot1}b), we present cuts for three potential surfaces $V_{1}(z)$ and different values for the pseudomomentum $K$. The magnetic field strength is again set to $B=1.41\, \rm T$, which gives a critical pseudomomentum of $K_{\rm crit} = 0.03\hbar/a_0$. In Fig.\ (\ref{plot1}b), the blue solid line represents a pseudomomentum of $K=3K_{\rm crit}$. Obviously, the curve $V_1(z)$ possesses a pronounced potential minimum at $z_{\rm min} \approx -0.65\, \mu \rm m$ and a lowering of $18\, \rm meV$ with respect to the saddle point. In contrast, for critical pseudomomentum $K=K_{\rm crit}$ (green solid line) the minimum of $V_1(z)$ has turned into a plateau at around $z \approx 180\, \rm nm$. Finally, the red solid curve represents the situation for even smaller pseudomomenta, here for $K=K_{\rm crit}/3$. We clearly see that the plateau has vanished and the curve is monotonically increasing for $z \rightarrow -\infty$. 

In summary, one observes that both the external field strengths $\bfmath{E}$, $\bfmath{B}$ as well as the pseudomomentum $\bfmath{K}$ provide the possibility to specifically address the topological properties of the potential surfaces. However, as $\bfmath{K}$ is related to an effective electric field, it is sufficient to vary only the external fields to obtain a complete understanding of the potential surfaces. Therefore, we set $\bfmath{K}=0$ for the rest of this work and only address the strengths and orientation of the external field parameters.

\subsubsection{Inclusion of spin couplings}
Next, we include both the spin-orbit coupling $H_{\rm so}$ and the spin-field coupling $H_B$ into the external potential term. For this reason, we define the total giant-dipole potential 
\begin{eqnarray}
V_{\rm{gd}}(\bfmath{r})\equiv V(\bfmath{r})+H_{\rm so}+H_B.
\end{eqnarray} 
We go back to the coupled hole spin $\bfmath{J}$ and obtain
\begin{eqnarray}
\hspace{-1.5em}V_{\rm gd}(\bfmath{r})=V(\bfmath{r})+\frac{\bar{\Delta}}{3}\left(\bfmath{J}^2-\frac{3}{4} \right)-\frac{\bar{\mu}_B}{2}\left( \bfmath{J}+3\bfmath{S}_h \right)\cdot\bfmath{B}. \label{Vgd_com}
\end{eqnarray}
If we neglect the term $H_B$, the remaining giant-dipole potential is a bilinear expression with respect to the $J_i,S_{h,j}$ angular momentum components, i.e.\
\begin{eqnarray}
V_{\rm gd}(\bfmath{r})&=&\sum_{ij}\left( \alpha^{(1)}_{ij}J_{i}J_{j}+\alpha^{(2)}_{ij}J_{i}S_{h,j}+\alpha^{(3)}_{ij}S_{h,i}S_{h,j} \right),\nonumber \\  &\ &\alpha^{(k)}_{ij}\in \mathbb{R}\ \ \alpha^{(k)}_{ij}=\alpha^{(k)}_{ji},\ \ k=1,2,3. \label{gd_bilinear}
\end{eqnarray}
Obviously, $V_{\rm gd}$ is time-reversal symmetric. Because $\bfmath{J}$ is a half-integer angular momentum, the remaining potential surfaces are at least doubly degenerate which is a direct consequence of the Kramers degeneracy theorem \cite{Kramers1930}.
A simple analysis shows that the spin-orbit coupling dominates over the spin-field coupling term $H_B$ far below a critical field strength of $B \approx 2\Delta/3\mu_B \approx 1541\, \rm T$. 
Including both spin terms in Eq.\ (\ref{Vgd_com}), we see that the bilinearity of Eq.\ (\ref{gd_bilinear}) is broken due to the spin-field coupling. Therefore, the twofold degeneracy of the potential curves provided by $V(\bfmath{r})$ is lifted, and we obtain six distinct potential curves $V^{(i)}_{\rm gd}(\bfmath{r}),\ i=1,...,6$. 

In Fig.\ \ref{plot2}(a), we show a cut of the potential curves for $y=x=0$ and applied field strengths of $B=1.41\, \rm T$ and $E=1\, \rm kV/cm$. At first sight, we can distinguish three different curves labeled with Roman numbers I,II,III. Similar to the previous case of vanishing spin-orbit coupling all three potential surfaces possess saddle points at around $z_{\rm s}\approx -30\, \rm nm$. Additionally, the potentials I and II possess local minima around $z^{\rm I}_{\rm min} \approx -1.42\, \mu \rm m$ and $z^{\rm II}_{\rm min} \approx -1.59\, \mu \rm m$, respectively. 

In contrast, the potential curve III does not possess any local potential minima. Due to the spin-orbit coupling, the energies of the saddle points are shifted by an amount of $\Delta$ and the two minima of the curves I and II are separated by around $100\, \rm meV$. Although only three curves are visible in Fig.\ \ref{plot2}(a), these curves are almost degenerate as the spin-field coupling $H_B$ causes a splitting of the order of $\mu_B$. In particular, the surface I splits into the surfaces $V^{(1)}_{\rm gd}(\bfmath{r})$ and $V^{(2)}_{\rm gd}(\bfmath{r})$, while II and III split into $V^{(3,4)}_{\rm gd}$ and $V^{(5,6)}_{\rm gd}(\bfmath{r})$, respectively. 

To resolve this issue in more detail, we have zoomed into the potential curve II which is shown in the inset of Fig.\ \ref{plot2}(a). Here we clearly see the two distinct potential curves $V^{(3)}_{\rm gd}$ and $V^{(4)}_{\rm gd}$ separated be an energy of around $190\, \mu \rm eV$. It turns out that the splitting induced by $H_B$ does not change the topological properties of neighboring potential surfaces. Thus we can neglect the spin-field coupling $H_B$ for the rest of this work. In this case, the potential curves are doubly degenerate and we have $V^{(1)}_{\rm gd}=V^{(2)}_{\rm gd}$, $V^{(3)}_{\rm gd}=V^{(4)}_{\rm gd}$ and $V^{(5)}_{\rm gd}=V^{(6)}_{\rm gd}$. Therefore, we restrict the analysis of the potential curves to $V^{(1)}_{\rm gd}(\bfmath{r})$ and $V^{(3)}_{\rm gd}(\bfmath{r})$, respectively. Furthermore, we introduce two additional quantities, namely the potential depth $V^{(i)}_{\rm d}$ and the lowering $V^{(i)}_{\Delta}$ of the potential minima with respect to the spin-orbit energy level $\Delta$ defined as
\begin{eqnarray}
V^{(i)}_{\rm d}&=&\lim\limits_{x \rightarrow \infty} V^{(i)}_{\rm gd}(x,0,z^{(i)}_{\rm min}) -V^{(i)}_{\rm gd}(\bfmath{r}^{(i)}_{\rm min}),\nonumber \\
V^{(i)}_{\Delta}&=&\Delta-V^{(i)}_{\rm gd}(\bfmath{r}^{(i)}_{\rm min}),\ \ i=1,3,5.
\end{eqnarray}

With increasing field strengths, the potential $V(\bfmath{r})$ dominates the spin-orbit coupling. This is shown in Fig.\ \ref{plot2}(b) where we present the potential curves $V^{(1)}_{\rm gd}$ and $V^{(3)}_{\rm gd}$ together with the exact potential $V_{1}(\bfmath{r})$ from Eq.\ (\ref{Vgd_analytic}) for applied field strength of $B=1.41\, \rm T$ and $E=4\, \rm kV/cm$. We clearly see that the two potential curves are still separated by the order of $\Delta$ but $V_{1}(\bfmath{r})$ is a good approximation for $V^{(3)}_{\rm gd}(\bfmath{r})$. In summary, for fixed $B$ we observe the following general features of the potentials:
\begin{enumerate}
 \item [(1)] The potential surface $V^{(5)}_{\rm gd}$ does not possess any potential minimum but a single saddle point.
 \item [(2)] The minima positions $z^{(i)}_{\rm min},\ i=1,3$ decrease with increasing $E$, the outer wells move away from the Coulomb singularity along the negative $z$-axis.
 \item [(3)] The minima of the outer wells decrease with increasing $E$ but are bounded from above by the spin-orbit coupling $\Delta$, i.e.\ $V^{(i)}_{\Delta} \ge 0,\ i=1,3,5$.
 \item [(4)] The saddle point positions $z^{(i)}_{\rm s},\ i=1,3,5$ increase with increasing $E$, the saddle point maximum moves towards the Coulomb singularity.
\end{enumerate}

Furthermore, we see that within the range of the applied field strengths we can easily achieve regimes where the spacing $V^{(i)}_{\Delta},\ i=1,3$ of the potential wells exceeds the binding energies of the field-dressed excitonic species. In Fig.\ (\ref{plot2}b), we indicate this fact by comparing the potential spacing $V^{(3)}_{\Delta}$ with the ground state binding energy $\Delta E_{1S(y)}$ of the $1S$-exciton of the yellow series which is approximately $150\, \rm meV$. For the given field strengths in Fig.\ \ref{plot2}(b), we obtain $V^{(3)}_{\Delta} \approx 1\, \rm eV$, which is larger than the excitonic binding energies and the spin-orbit spacing $\Delta$. As a consequence, the spin-orbit coupling term $H_{\rm so}$ can be treated as a perturbation to the potential $V(\bfmath{r})$. Additionally, for such field strengths the excitonic states, which are mostly determined by the inner Coulomb potential, will couple to the giant-dipole continuum. This coupling possibly leads to a broadening of the excitonic spectral lines. At this point, we emphasize that for a precise study we have to compare the potential spacing $V^{(i)}_{\Delta}$ with the binding energies of the field-dressed excitonic species as they have been analyzed in recent experimental studies \cite{Schweiner2016}. However, in this particular work it was shown that for magnetic field strengths up to $3\, \rm T$ the energetic shift due to the external magnetic field is of the order of several hundreds of $\mu \rm eV$ compared to the field-free ground state energy. These values are far below the potential splittings observed for the giant-dipole potential curves. 

\begin{figure}[h]
\centering
\begin{minipage}{0.45\textwidth} 
\includegraphics[width=0.9\textwidth]{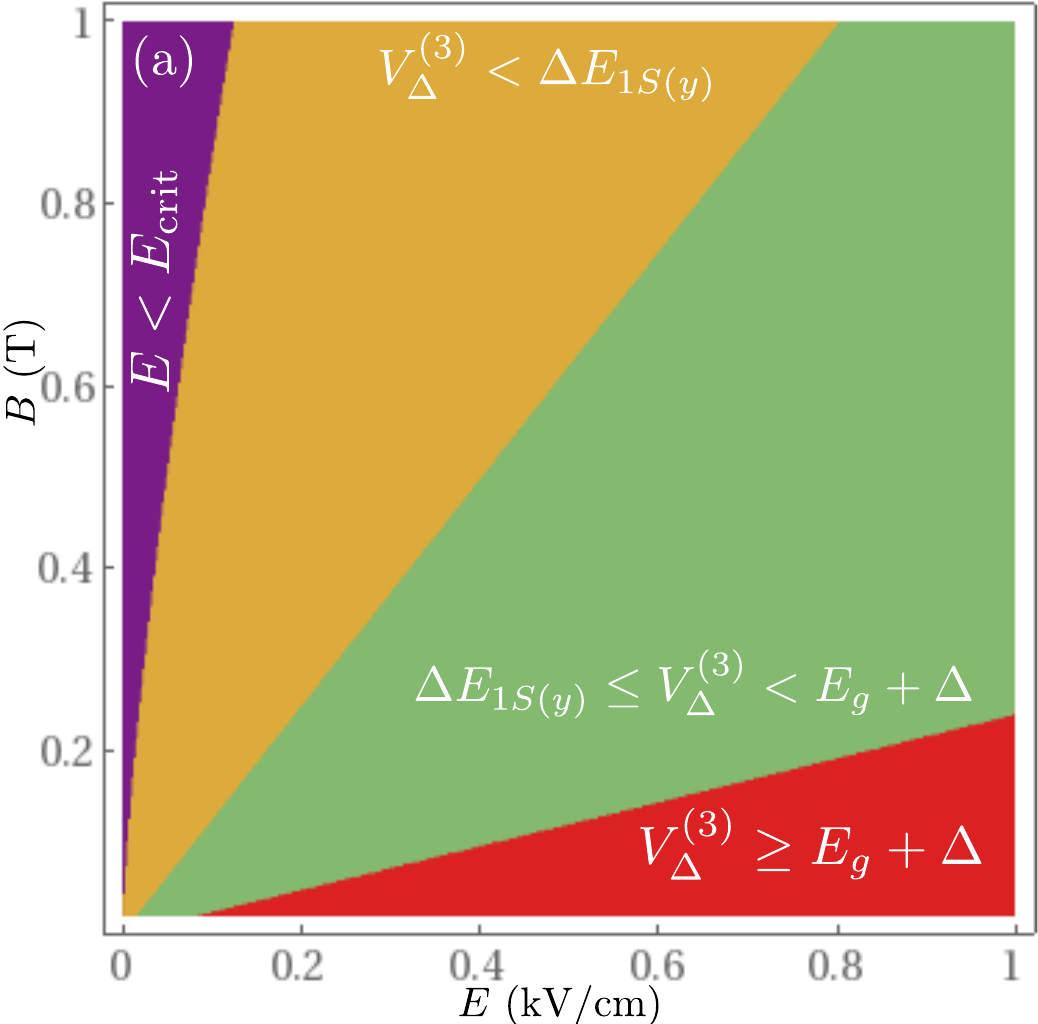}
\end{minipage}
\begin{minipage}{0.475\textwidth} 
\includegraphics[width=\textwidth]{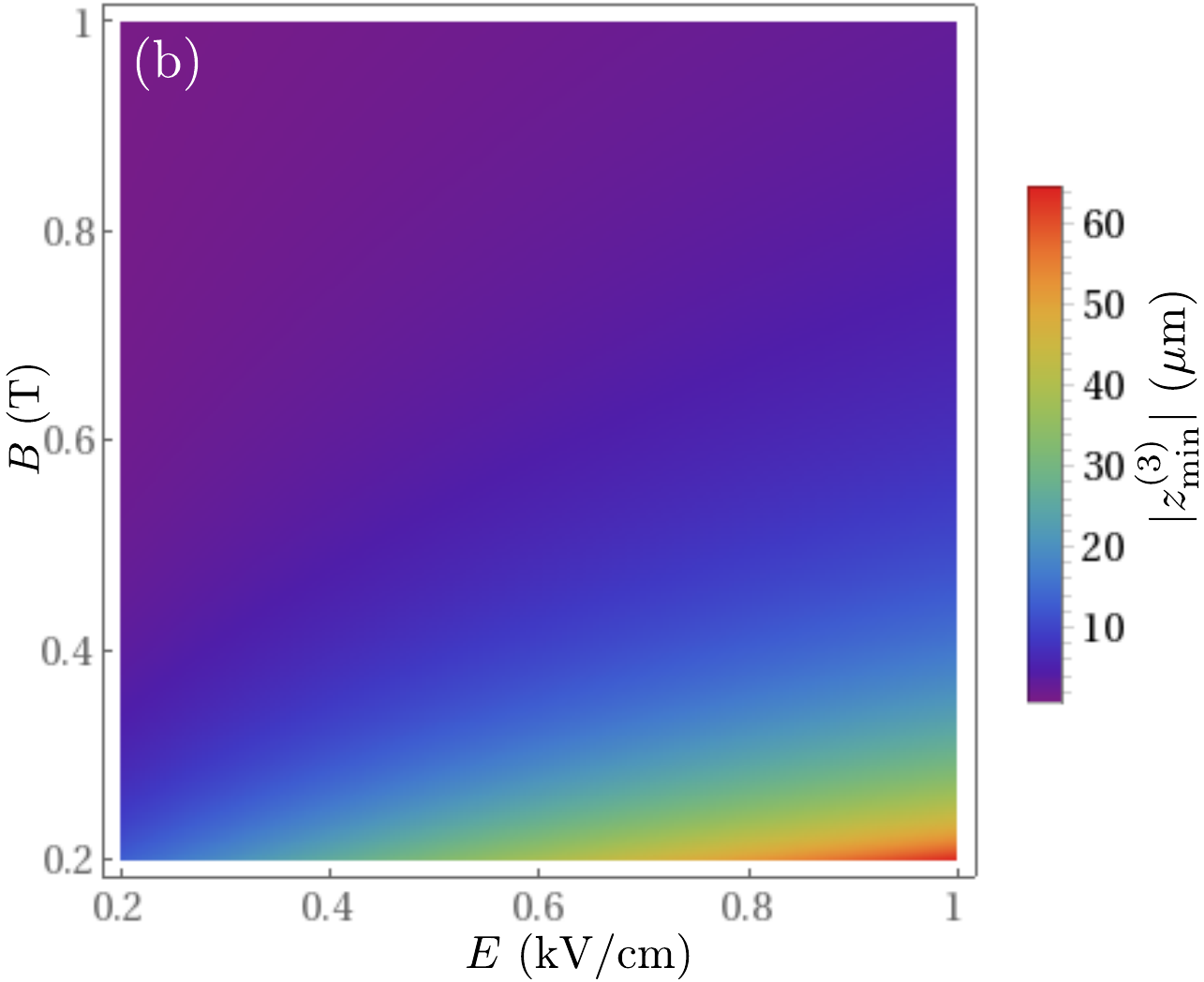}
\end{minipage}
\caption{(a) Phase-like diagram for potential curve $V^{(3)}_{\rm gd}$ with respect to the $1S$-yellow excitonic ground state. (b) The minimum position $|z^{(3)}_{\rm min}|$ as function of the external field parameters. Depending on the applied field strengths we obtain spatial separations in the range of $1-60\, \mu \rm m$.}
\label{phases_1S_1P}
\end{figure}

A more detailed study of the different addressable regimes is presented in Fig.\ \ref{phases_1S_1P}(a). There we show a phase-like diagram where the spacing $V^{(3)}_{\Delta}$ is compared to the binding energy of the $1S$-yellow exciton for electric and magnetic field strengths in the range between $0 \le E \le 1\, \rm kV/cm$ and $20\, {\rm mT} \le B \le 1\, \rm T$, respectively. In Fig.\ \ref{phases_1S_1P}(a) different colors (blue, yellow, green, red) illustrate different regimes. For instance, in case of electric fields below the critical value  $E_{\rm crit}$ the potential surface $V^{(3)}_{\rm gd}$ does not exhibit any potential well. This regime is indicated by a dark blue color. For sufficiently strong electric fields we can distinguish three additional regimes. In the yellow regime, $V^{(3)}_{\Delta}$ is less than the $1S$-yellow excitonic binding energy. In this case, bound excitonic states in the outer potential well are energetically still above the (field-dressed) $1S$-yellow excitonic ground state. This situation is changed for increasing electric field strength and indicated by the green sector in the diagram shown in Fig.\ \ref{phases_1S_1P}(a). Here, the electric field is so strong that $V^{\rm (3)}_{\Delta}$ is larger than the $1S$-yellow binding energy. As we have already mentioned, in this situation we expect line broadening of excitonic levels due to coupling to the giant-dipole continuum.

If we further increase the electric field strength we are able to reach a regime where the spacing $V^{\rm (3)}_{\Delta}$ even exceeds the band gap of the $\Gamma^{8+}$-valence and $\Gamma^{6+}$-conduction band, i.e.\ $V^{(3)}_{\Delta} > E_g+\Delta$ (dark red color). Obviously, this particular regime has to be taken with caution as the simple description of the exciton breaks down and more complicated theoretical approaches describing many-body interactions between the excitonic constituents and the electrons in the $\Gamma^{8+}$-valence band are required. In Fig.\ \ref{phases_1S_1P}(a), we see that for weak magnetic fields ($B \approx 50\, \rm mT$) and comparably low electric fields $E \approx 150\, \rm V/cm$ we easily address these extreme excitonic states. This behavior is reasonable as for low magnetic fields already minor electric fields can lead to a pronounced field-induced level shift of the excitonic energy, leading both to a pronounced Stark shift and the creation of an outer potential well. Because $\Delta/E_g \ll 1$, the minimum positions of these wells are well approximated by Eq.\ (\ref{positions}), which directly gives $|z^{(i)}_{\rm min}| \sim E/B^2$. This fact is analyzed in Fig.\ \ref{phases_1S_1P}(b) in more detail. Here we present the absolute value of the outer well position $|z^{(3)}_{\rm min}|$ as a function of the applied field strengths $B$ and $E$, respectively. We see that, depending on the applied field strengths, we obtain electron-hole separations in the range of $1-60\, \mu \rm m$. 
\begin{figure}[h]
\centering
\includegraphics[width=0.475\textwidth]{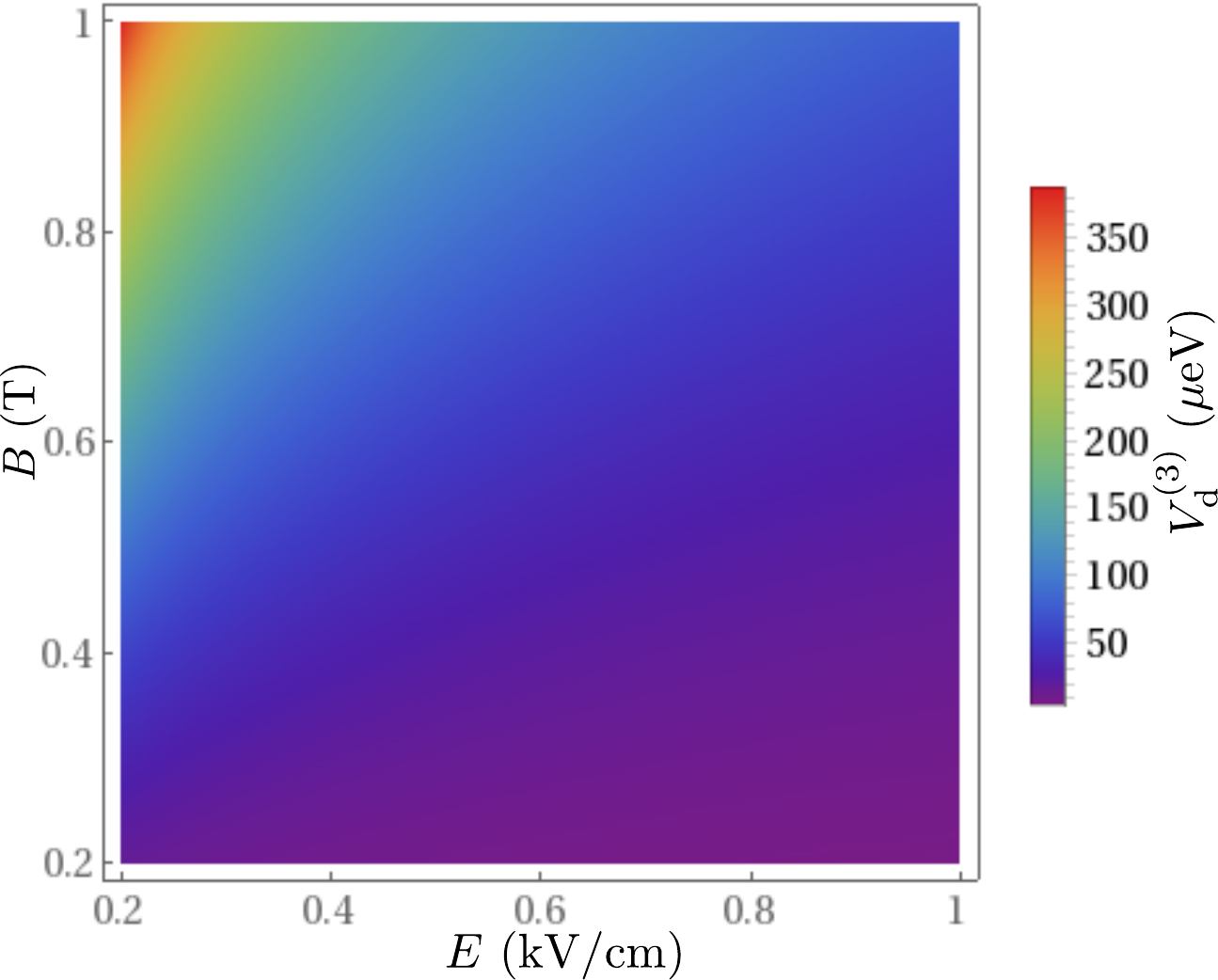}  
\caption{Potential depths $V^{(3)}_{\rm d}$ as a function of the applied field strengths. We obtain depths in the range of $5 - 380\, \mu \textrm{eV}$.}
\label{potdepth}
\end{figure}

In Fig.\ \ref{potdepth} we present the potential depth $V^{(3)}_{\rm d}$ for the potential surface $V^{(3)}_{\rm gd}$ as a function of electric and magnetic field strengths, respectively. For the applied field strengths $(0.2\, \textrm{kV/cm} \le E \le 1\, \textrm{kV/cm},\ 0.2\, \textrm{T} \le B \le 1\, \rm T)$ we obtain potential depths in the range of $5 - 380\, \mu \textrm{eV}$. 

At this point, we indicate that for sufficiently deep potential wells one can expect excitonic bound states in the outer wells. To obtain their full spectrum, one has to perform an exact diagonalization including the non-trivial kinetic energy terms discussed in Sec.\ \ref{gd_Hamiltonian}. These excitonic species are of completely different nature as the field-dressed excitons which are localized in the inner Coulomb-dominated region. As the excitons in the outer potential well possess a large spatial electron-hole separation, their electric dipole moment is expected to be exceptionally large. In particular, it can be approximated to be $d_{\rm ex}=|z^{(i)}_{\rm min}|$, which gives huge dipole moments in the range of $5 \cdot 10^5-2.8 \cdot 10^6\, \rm D$. Analogous to the atomic giant-dipole states, we might denote these kind of exotic excitonic states as giant-dipole excitons.

As it has been shown in previous works the binding energies of excitons in $\textrm{Cu}_2\textrm{O}$ reveal a slight deviation from a pure Rydberg series \cite{Scheel2014}. This deviation can be incorporated by employing the concept of quantum defects such that the excitonic binding energies are given by $\epsilon_{\rm ex}=-\mathcal{H}_{\rm ex}/(2(n-\delta^{(l)}_{n}))^2,\ \delta^{(l)}_{n} >0$ \cite{Schoene2016}. For $\rm S$-excitons of the yellow series we have $\delta^{(0)}_{n\ge10}\approx 0.5$.
Therefore, we easily determine the principal quantum number $n$ for which the binding energy of the Rydberg exciton is comparable to the potential depth to be in the range of $n=15$ ($V^{(3)}_{\rm d}=380\, \mu \rm eV$) up to $n=42$ ($V^{(3)}_{\rm d}=50\, \mu \rm eV$). At this point, we note that for more precise results one has to compare the binding energy of the excitonic giant-dipole ground states localized in the outer potential well with the binding energies of the field-dressed states localized in the inner region. However, for such an analysis further information regarding excitonic Landau-levels is required which is a topic of ongoing research. Finally, we evaluate the principal quantum number $n$ which is required to obtain the same binding energy as the giant-dipole state for the hydrogenic system as it has been analyzed for the field strengths applied in Ref.\ \cite{Dippel1994}. Most importantly, in the hydrogen atom the Rydberg constant is given as $\textrm{Ry}=13.6\, \rm eV$, which is around $160$ times higher than its excitonic counterpart. In Ref.\ \cite{Dippel1994}, the giant-dipole binding energy for a hydrogen atom has been determined to be approximately $274\, \mu \rm eV$. Together with the higher Rydberg energy this leads to principal quantum numbers of $n \approx 223$.\\
\subsection{Comparison of different magnetic field configurations}
At last, we consider a different field orientation of the magnetic field. In contrast to the giant-dipole species studied by Dippel {\it et\ al.} \cite{Dippel1994}, in the present study the quantization axis is determined by the symmetry properties of the $\textrm{Cu}_2\textrm{O}$ crystal. For this reason, we expect the giant-dipole potential surfaces to be explicitly dependent on the applied field orientation. To study this in more detail, we have fixed the electric field configuration to be parallel to the $z$-axis ($\bfmath{E} || [0 0 1]$) and chosen two distinct orientations for the magnetic field, namely $\bfmath{B} ||[1 0 0]$ and $\bfmath{B} || [ 1 1 0]$. 

More precisely, in Fig.\ \ref{multiplot}(a)-(d) we show two-dimensional potential surfaces $V^{(1)}_{\rm gd}(0,y,z) \equiv V^{(1)}_{\rm gd}(y,z)$ and $V^{(3)}_{\rm gd}(y,z)$ for two different field configurations for fixed magnetic field strengths $B=1.41\, \rm T$ and $E=1\, \rm kV/cm$, respectively. The figures Fig.\ 6(a,b) show the potential surfaces with the magnetic field oriented along the $[100]$ direction. For both surfaces we clearly see two distinct potential minima at $z^{(1)}_{\rm min} \approx -1.3\, \mu \rm m,\ y^{(1)}_{\rm min}=0$ and $z^{(3)}_{\rm min} \approx -1.31\, \mu \rm m,\ y^{\rm (3)}_{\rm min}=0$, respectively. Both potentials monotonically increase for $z \rightarrow -\infty$ and $|y|\rightarrow \infty$, which simply reflects the properties of the diamagnetic field term. For $(y,z) \rightarrow 0$ the Coulomb interaction becomes the dominant part in the electron-hole interaction potential. For this reason both potential curves possess a singularity near the origin. 

For comparison, in Fig.\ \ref{multiplot}(c,d) we show the potential surface $V^{(1)}_{\rm gd}(y,z)$ and $V^{(3)}_{\rm gd}(y,z)$ for fixed field strengths $B=1.41\, \rm T$, $E=1\, \rm kV/cm$, but now for a magnetic field oriented along the $[110]$ direction. In contrast to the previous field orientation, we now plot the potential surfaces not in the $yz$-plane, but in the plane spanned by the vectors $\bfmath{e}_z$ and $\bfmath{e}_{\eta}=(\bfmath{e}_y-\bfmath{e}_x)/\sqrt{2}$, respectively. Here, we label the coordinate with respect to $\bfmath{e}_{\eta}$ with $\eta$. Similar to the first field configuration we find potential surfaces possessing localized minima on the negative $z$-axis. However, now the explicit positions of these minima have changed and are given by $z^{\rm (1)}_{\rm min} \approx -0.4\, \mu \rm m,\ \eta^{\rm (1)}_{\rm min }=0$ and $z^{\rm (3)}_{\rm min} \approx -1.45\, \mu \rm m,\ \eta^{\rm (3)}_{\rm min }=0$, respectively. Furthermore, compared to the previous field configuration the global topologies have changed as the potential surfaces are increasing faster for $|\eta|\rightarrow \infty$ and $z \rightarrow -\infty$. From this analysis we clearly see that the topologies of the giant-dipole potential surfaces depend on the explicit configuration of both the external electric and magnetic field configurations. 

In summary, we have analyzed the topological properties of the potential energy surfaces in much detail. However, the question still remains if these potential wells provide the possibility of bound states. Although this question is still a topic of ongoing research, we can deduce some information making some estimations analogously to the study performed by Dippel {\it et.\ al} \cite{Dippel1994} for atomic systems. In this work it was shown that the energy spacing of bound giant-dipole states is not entirely determined by the trapping frequencies of the harmonic approximation near the potential minimum, but both the sum and differences of these frequencies. This leads to a comparable small energy spacing in case that at least two frequencies are of nearly the same magnitude, providing a dense spectrum of bound states within a specific potential well. As we can address the potential's topologies directly via both the field strength and orientations we have direct access to the trapping frequencies. Thus, we feel confident to tune the external field parameters in such a way that bound excitonic giant-dipole states are present.
\begin{widetext}
\begin{minipage}[t]{0.5\textwidth} 
\includegraphics[width=0.8\textwidth]{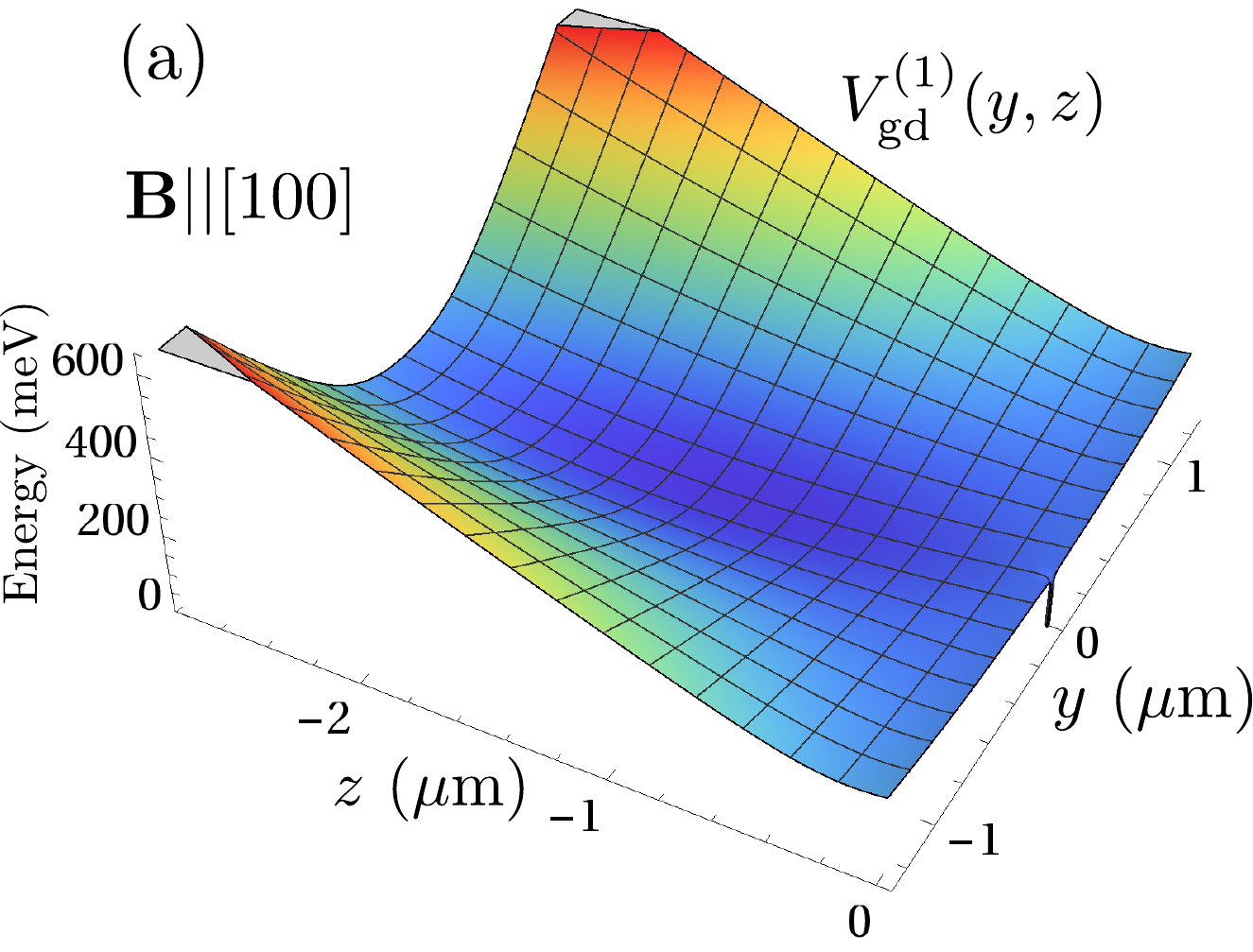}  
\end{minipage}
\begin{minipage}[t]{0.5\textwidth} 
\includegraphics[width=0.8\textwidth]{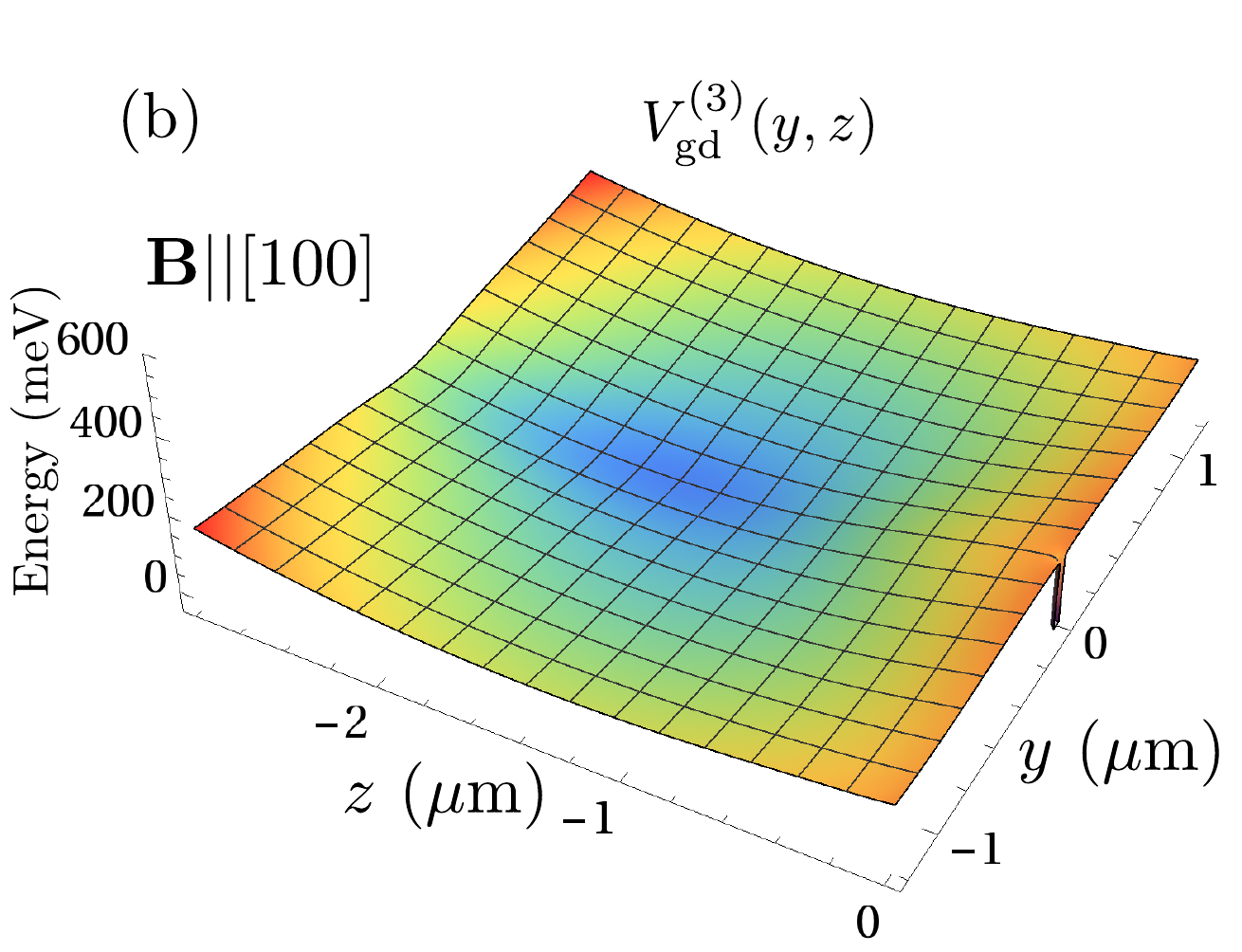} 
\end{minipage}
\begin{minipage}{0.5\textwidth} 
\includegraphics[width=0.8\textwidth]{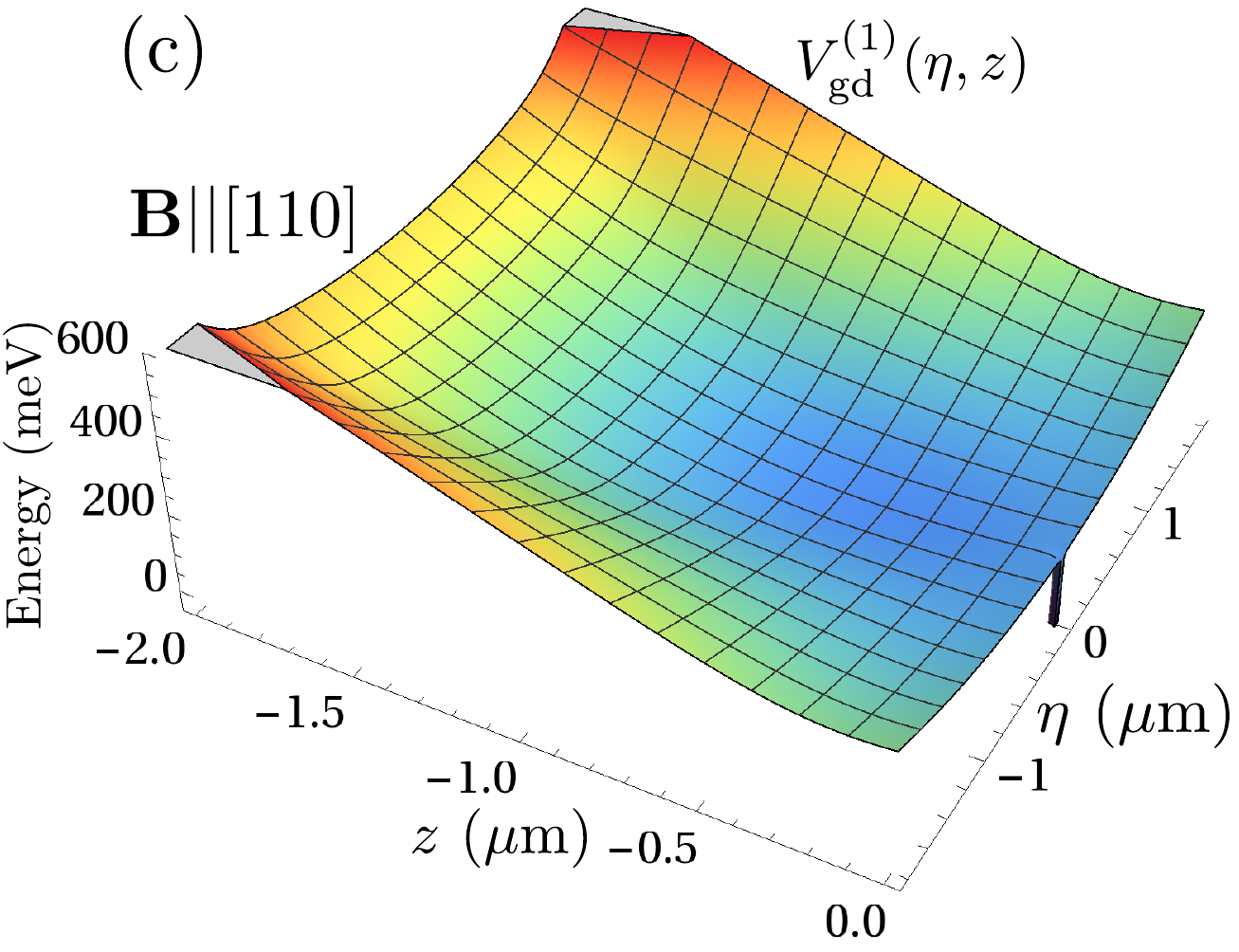}  
\end{minipage}
\begin{minipage}{0.5\textwidth} 
\includegraphics[width=0.78\textwidth]{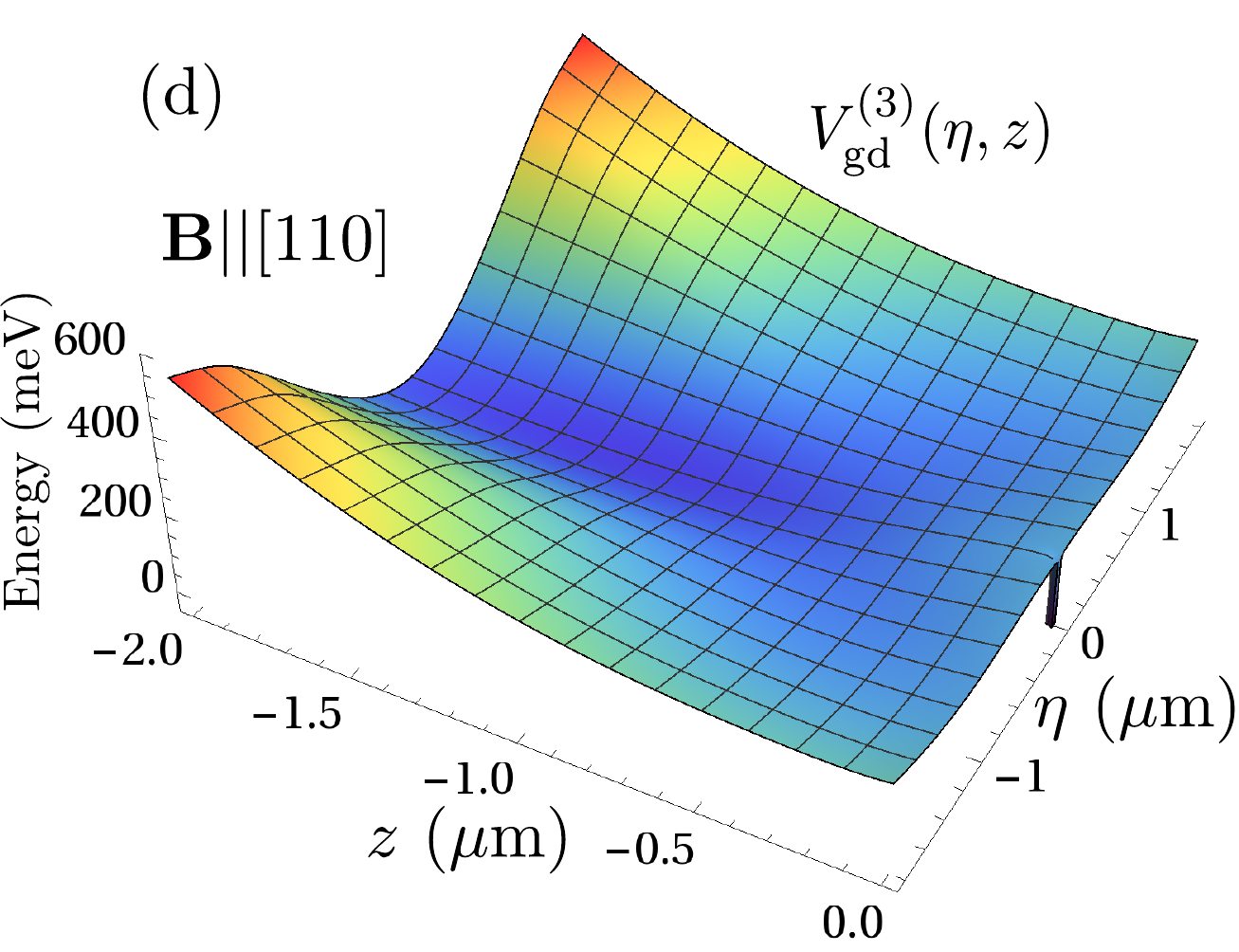} 
\end{minipage}
\begin{figure}[H]
\caption{Two-dimensional potential energy surfaces $V^{(1)}_{\rm gd}(y/\textcolor{blue}{(\eta)},z)$, $V^{(3)}_{\rm gd}(y/\textcolor{blue}{(\eta)},z)$ for two distinct magnetic field configurations ($B=1.41\, \rm T$, $E=1\, \rm kV/cm$). Figures (a,b) show $V^{(1,3)}_{\rm gd}$ for the magnetic field oriented along the $[1 0 0]$ direction. Figures (c,d) show $V^{(1,3)}_{\rm gd}$ for the magnetic field oriented along the $[1 1 0]$ direction. All plots show potential surfaces perpendicular to the applied magnetic field and $x=0$. The global topologies of the potential surfaces clearly depend on the specific field configuration. In all figures the electric field is oriented along the $[001]$ direction.}
\label{multiplot}
\end{figure}
\end{widetext}
\section{Summary and conclusions \label{conclusion}}
In the present paper, we have provided the theory of excitons in $\textrm{Cu}_2\textrm{O}$ subject to crossed electric and magnetic fields. In particular, we have performed a gauge-independent pseudoseparation of the center-of-mass motion for the electron-hole system in crossed fields. In the resulting Hamiltonian we were able to identify terms that depend on a gauge-dependent vector potential and belong to the kinetic energy of the relative motion. The effective gauge fields indicate possible non-Abelian gauge-field description. The complementary terms of the Hamiltonian are gauge-independent and can therefore be assigned to an effective single-particle interaction potential for the relative motion. From this interaction potential we were able to present a number of potential surfaces for the relative electron-hole dynamics. 

Due to the coupling of the center-of-mass motion to the internal degrees of freedom, the effective electron-hole potential exhibits outer potential wells with depths up to $380\, \mu \rm eV$. This leads to large spatial electron-hole separation in the range of $1-60\, \mu \rm m$. Due to this large distance of the well from the Coulomb singularity, bound excitonic states in the outer well potentially possess large permanent electric dipole moments. We showed that within the range of standard laboratory field strengths for both the electric ($E \le 1\, \rm kV/cm)$ and magnetic fields ($B \le 1\, \rm T$), one can easily address the topological properties of the potential surfaces by changing the applied field strengths and field orientations, respectively. 

Furthermore, we have shown that in $\textrm{Cu}_2\textrm{O}$ it is sufficient to excite the excitons to principal quantum numbers starting from $n \approx 10$ in order to obtain binding energies comparable to the depth of the excitonic potential surfaces. In case of hydrogen one would have to excite the electron into Rydberg states with extremely high $n \ge 223$. For this reason, it is obvious that excitonic systems are a much more promising candidate for possible experimental realizations of giant-dipole species.

For an experimental verification of the existence of excitonic giant-dipole states there are distinct possible routes. One would be the spectroscopic observation of state-to-state transitions. For this, a precise knowledge of the excitonic level spacing is required. To obtain the giant-dipole spectrum, a complete analysis of the excitonic Hamiltonian including the non-trivial kinetic energy terms has to be performed. This issue is a topic of ongoing research. 

Another approach to experimental verification is the direct measurement of the electric dipole moment. To estimate the electric  dipole moment for excitonic giant-dipole states we approximate the distance between the electron and the hole by the distance between the minimum of the outer potential well and the Coulomb singularity. Applying this approximation, the dipole moment can reach $d\approx 3\cdot 10^6\, \rm D$. For comparison, the dipole moment of excitons confined to individual self-assembled ring-shaped quantum dots in the insulator region of a metal-insulator-semiconductor heterostructure have been determined to be around $150\, \rm D$ \cite{Warburton2002}, which is around four orders of magnitude less the predicted dipole moments of the novel excitonic states. 

Finally, we point out that in the case of excitonic giant-dipole states the present study took place in a complex solid-state environment, in contrast to ultracold atomic species, where the experimental preparation in various trap geometries provides a much higher degree of external control \cite{Cumbescot1985,Kushwaha2001}. For instance, the application of external electric fields leads to a strong response of the material such as polarization and shielding effects of electrons in the conduction bands forming a quasi-free electron plasma. Therefore, it is not a priori clear which conditions are realized inside the considered solid-state system in case external parameters are applied \cite{Lewenstein2007}. Thus, excitonic giant-dipole states provide a plethora of interesting problems which can be addressed in future studies.
\begin{acknowledgments}
We thank Prof.\ H.\ Stolz and M.\ Sc.\ Florian Sch\"one for helpful and fruitful discussions about semiconductors and excitonic physics. Furthermore, we thank Dr.\ Dirk Semkat for cross reading  the manuscript. We acknowledge support by the Focus Programme SPP 1929 GiRyd by the Deutsche Forschungsgemeinschaft (DFG).
\end{acknowledgments}
\begin{widetext}
\section*{Appendix: Giant-Dipole Potential \label{appendix}}
The coefficients $\Omega_i,\ i=1,2,3$ are functions of the Luttinger parameters $\gamma_i$ and the constants $C_{i}$ \cite{Luttinger1954}. In particular, they are
\begin{eqnarray*}
\Omega_1&=&\frac{1}{2m_e}-\frac{C_1}{m_e}+\frac{1}{18}\left(9C^{2}_{1}+2C^{2}_{2}+3C^{2}_{3} \right)-\frac{\gamma_2}{18\gamma^{'}_{1}}\left( 24C_1C_2-4C^{2}_{2}-3C^{2}_{3} \right) \\
&\ &-\frac{\gamma_3}{12 \gamma^{'}_{1}}C_3(24C_1-4C_2-3C_3), \\
\Omega_2&=&\frac{C_2}{3m_e}-\frac{1}{72}(24C_1C_2-4C^{2}_{2}-3C^{2}_{3})+\frac{\gamma_2}{3\gamma^{'}_{1}}(3C^{2}_{1}-2C_1C_2+C^{2}_{2}-C^{2}_{3})\\
&\ &-\frac{\gamma_3}{24\gamma^{'}_{1}}C_3(12C_1-2C_2+3C_3),\\
\Omega_3&=&\frac{C_5}{m_e}-\frac{C_5}{24}(24C_1-4C_2-3C_3)-\frac{\gamma_2}{12\gamma^{'}_{1}}(12C_1C_3-2C_2C_3+3C^{2}_{3})\\
&\ &+\frac{\gamma_3}{24\gamma^{'}_{1}}(72C^{2}_{1}-24C_1C_2-36C_1C_3-16C^{2}_{2}-12C_2C_3+27C^{2}_{3}).
\end{eqnarray*}
with $m_e=0.985\gamma^{'}_{1}$. The matrices of the quasi-spin $I=1$ are defined as in Ref.\ \cite{Luttinger1956} as
\begin{eqnarray*}
I_k=-i\sum_{lm} \varepsilon_{klm}(\bfmath{e}_l \otimes \bfmath{e}_m). 
\end{eqnarray*}
Then, the matrix representation of the giant-dipole potential $V(\bfmath{r})$ from Eq.\ (\ref{gd_V}) is given by
\begin{eqnarray*}
V(\bfmath{r})=
\begin{bmatrix}
V+\Omega_2 (3\tilde{K}^{2}_{1}-\tilde{K}^2) & \Omega_3 \tilde{K}_{1}\tilde{K}_{2} & \Omega_3 \tilde{K}_{1}\tilde{K}_{3} \\
\Omega_3 \tilde{K}_{1}\tilde{K}_{2} & V+\Omega_2 (3\tilde{K}^{2}_{2}-\tilde{K}^2) & \Omega_3 \tilde{K}_{2}\tilde{K}_{3}\\
\Omega_3 \tilde{K}_{1}\tilde{K}_{3} & \Omega_3\tilde{K}_{2}\tilde{K}_{3}& V+\Omega_2 (3\tilde{K}^{2}_{3}-\tilde{K}^2)
\end{bmatrix}
\end{eqnarray*}
with $V=\Omega_1 \tilde{K}^2+\bfmath{E}\cdot\bfmath{r}-\frac{1}{r}$, $\tilde{\bfmath{K}}=\bfmath{K}+\bfmath{B}\times \bfmath{r}$.
\end{widetext}
\section*{Bibliography}
\end{document}